\documentclass{article}
\usepackage[utf8]{inputenc}
\usepackage{amsmath}
\usepackage{amssymb}
\usepackage{amsthm}
\usepackage{authblk}
\usepackage{hyperref,braket}
\usepackage{cleveref} %always load 
\usepackage{xcolor,graphicx}

\newcommand{\bq}{\begin{equation}} \newcommand{\eq}{\end{equation}}
\newcommand{\bqali}{\begin{equation}\begin{aligned}}
\newcommand{\eqali}{\end{aligned}\end{equation}}
\newcommand\D{\operatorname{d}\!}
\renewcommand\k{{\bf k}}
\renewcommand\r{{\bf r}}
\newcommand\p{{\bf p}}
\newcommand\q{{\bf q}}
\newcommand\z{{\bf z}}

\newcommand\x{{\bf x}}
\newcommand\y{{\bf y}}
\newcommand\g{{\bf g}}
\newcommand\kB{k_\text{\tiny B}}
\newcommand{\rC}{r_\text{\tiny C}}

\newcommand{\erf}{\operatorname{erf}}

\title{Spontaneous Collapse Models}

\author[1,2]{Matteo Carlesso}
\author[2]{Sandro Donadi}
\affil[1]{Department of Physics, University of Trieste, Strada Costiera 11, 34151 Trieste, Italy}
\affil[2]{Centre for Quantum Materials and Technologies,
School of Mathematics and Physics, Queens University, Belfast BT7 1NN, United Kingdom}

\date{\today}

\begin{document}

\maketitle

\section{Abstract}
Collapse models are phenomenological models introduced to solve the measurement problem in quantum mechanics. They modify the Schr\"odinger equation by adding non-linear and stochastic terms, which induce the wavefunction collapse in space. The collapse effects are negligible for microscopic systems but become dominant in the macroscopic regime, thus also describing coherently the quantum-to-classical transition. Collapse models make different predictions compared to those of quantum mechanics; hence they can be tested. Here we introduce the most relevant collapse models present in the literature, and describe their main features. We also discuss how one can test them in different experiments, underlying the differences with predictions of quantum mechanics, and show how these experiments can set bounds on the collapse parameters. We conclude with a brief summary of the colored  and dissipative  generalization of such models and their experimental tests. 

{\bf Keywords}: quantum measurement problem, modifications quantum mechanics, collapse models, GRW model, CSL model, Di\'osi-Penrose model, experimental tests of quantum mechanics

\tableofcontents

\section{Introduction}

Quantum mechanics can be safely considered one of the most successful theories devised so far.  It has completely changed our understanding of nature, from the structure and interaction between known constituents of matter to the prediction of new particles, and  --- going beyond the blue-sky perspective alone --- it gave birth 
to new revolutionary technologies we employ on a daily basis.
Despite its undoubted success, 
the domain of its fundamental principle, being the quantum superposition principle, is well limited to microscopic and mesoscopic realms. 
According to this principle, {one can prepare a quantum system to be simultaneously in distinguishable physical configurations (such as ``here'' and ``there'')}. This is the basis for several technological applications of quantum mechanics.  
However, when someone tries to extend the domain of such a principle to the macroscopic world, they will  --- without exception --- encounter a puzzling situation. Although quantum theory does not set  any clear limit on the validity of the superposition principle, thus providing the impression that is universally valid, we do not experience any superposition at the macroscopic scale. One possible explanation is that 
the quantum superposition principle breaks down if the system exceeds a certain scale of dimension or complexity when moving from the microscopic to the macroscopic regime. Importantly, such a breakdown would also imply that quantum mechanics is not a universal theory but only a suitable limit of a more general one, yet to be discovered.

With the purpose of finding a solution for such a puzzle, without  putting in place a more fundamental theory, but taking a more phenomenological approach, collapse models were proposed. {These models modify} quantum mechanics providing a consistent framework where the wavefunction of a system collapses spontaneously (in space) and thus {explain} the breakdown of the superposition principle. As we will see in detail below, thanks to an in-built amplification mechanism, the collapse effects are negligible for microscopic systems but become considerable for macroscopic ones. Indeed, one wants  microscopic systems to continue to follow the rules of quantum mechanics, which are well-verified experimentally with no visible effect of a collapse dynamics. Conversely, one wants a theory that makes the wavefunctions of macroscopic systems well localized in space, and not superpositions, thus 
reconciling the theory with our everyday experience.

It is fundamental to underline that collapse models are not one among the possible interpretations of standard quantum mechanics: they fundamentally modify the latter and such modifications provide testable predictions, which differ from those of quantum mechanics. This is a unique feature of these models in contrast to several alternatives. We will dwell more on this in Sec.~\ref{sec_exp}.

\subsection{Problem at hand}\label{hand}
The debate on the interpretation of quantum mechanics started right after the formulation of the theory, and it is far from being resolved. Its culprit revolves around the so-called quantum measurement problem, also known as the macro-objectification problem.
According to quantum mechanics, one has the following dynamical postulate:
\paragraph{{\bf Dynamical postulate 1:}} Systems evolve according to the Schr\"odinger equation, which reads
\begin{equation}\label{sch}
    i\hbar\frac{\D}{\D t}|\psi(t)\rangle=\hat H|\psi(t)\rangle,
\end{equation}
where $\hbar=h/2\pi$ is the reduced Planck constant, $\ket{\psi(t)}$ is the wavefunction of the system describing its state at time $t$, and $\hat H$ is its Hamiltonian.
One of the main properties of the Schr\"odinger equation is its linearity. It follows that given two states $\ket{\psi_a(t)}$ and $\ket{\psi_b(t)}$ that are solutions of Eq.~\eqref{sch}, any superposition of these states, i.e. $|\psi(t)\rangle=a|\psi_{a}(t)\rangle+b|\psi_{b}(t)\rangle$ with $|a|^2+|b|^2=1$, is also a solution. In particular, Eq.~\eqref{sch} preserves superpositions in time.

The Schr\"odinger equation is also valid  for multi-particle systems and the theory does not set any limit to the number of particles. In principle, this allows for the generation of superpositions of arbitrary large systems starting from those of microscopic ones. For example, imagine a particle whose spin state is an equal superposition of the spin pointing towards up and towards down, i.e.~$|\psi\rangle=(\ket{\uparrow}+\ket{\downarrow})/\sqrt{2}$. We assume that such a particle interacts with a second one, whose state is $\ket{\phi}$,  in such a way that  the common state $\ket{\varPsi}$ evolves according to
\begin{equation}
|\varPsi\rangle=|\psi\rangle|\phi\rangle\longrightarrow|\varPsi(t)\rangle=\hat U(t)\ket{\varPsi}=\frac{|\uparrow\rangle|\phi^{\uparrow}\rangle+|\downarrow\rangle|\phi^{\downarrow}\rangle}{\sqrt{2}},
\end{equation}
where $\hat U(t)=e^{-i\hat H t/ \hbar}$ evolves the state according to the unitary dynamics due to the Schr\"odinger equation.
Instead of considering only two particles, one can imagine doing this with $N+1$, then the state after the interaction is
\begin{equation}\label{macrosup}
|\varPsi(t)\rangle=\frac{|\uparrow\rangle|\phi_{1}^{\uparrow}\rangle\dots|\phi_{N}^{\uparrow}\rangle+|\downarrow\rangle|\phi_{1}^{\downarrow}\rangle\dots|\phi_{N}^{\downarrow}\rangle}{\sqrt{2}}.
\end{equation}
This is essentially what happens in the famous thought experiment of the Schr\"odinger cat: a system prepared in a quantum superposition interacts with another one, or several ones, and the common final state of the system becomes entangled. If $\ket{\phi_1}\dots\ket{\phi_N}$ describes the state of the $N$ particles constituting a single macroscopic object, then the latter is --- due to the interaction with the initially superposed particle --- in a quantum superposition. However, such superpositions are not observed. To solve such an inconsistency, a second dynamical postulate is introduced in the theory. 

\paragraph{{\bf Dynamical postulate 2:}}  During the measurement process of a system, a measurement device collapses the state $\ket{\psi(t)}$ of the system in one of the eigenstates $\ket{o_n}$ of the operator $\hat O$ associated to the measured observable $\mathcal O$, with a probability given by the Born rule $P[o_n]=|\braket{o_n|\psi(t)}|^2$. Namely, the state changes as
\begin{equation}
    \ket{\psi(t)}\ \text{before measurement}\longrightarrow\ket{o_n}\ \text{after measurement}.
\end{equation}
This is the famous wavepacket reduction (or collapse) postulate. The corresponding dynamics is drastically different from that given by the evolution in postulate 1: the Schr\"odinger dynamics is fully linear (thus it preserves superpositions) and deterministic, while the collapse is non-linear and stochastic. In particular, the non-linearity is given by how the probabilities $P[o_n]$ are computed.

Having two different dynamical principles within the same theory is not by itself a fundamental problem, although it is unique to quantum theory. The real issue is that postulate 2 is fundamentally ambiguous in describing the collapse process. Indeed, there is no rule that determines which systems are eligible to be considered as measurement devices or observers --- and thus, when involved, lead to the wavefunction collapse ---  and which are instead systems that can be only observed. Indeed, all observers and measurement devices are made by the same microscopic constituents, which in turn should follow the Schr\"odinger equation. Such ambiguity is essentially the measurement or macro-objectification problem.

It has been advocated that decoherence resolves the measurement problem   \cite{anderson2001science}. This is not correct, even though decoherence plays an important role in understanding some aspects of the so-called quantum-to-classical transition, namely the mechanism due to which systems stop behaving as quantum and start to follow classical behaviours. 
To understand the action of decoherence, we introduce the concept of a statistical operator $\hat \rho$ as
\begin{equation}
    \hat \rho=\sum_{jk}p_{jk}\ket{\psi_j}\bra{\psi_k},
\end{equation}
where $\{\ket{\psi_j}\}_j$ is a suitable basis of the system's associated Hilbert space, and $p_{jk}=\braket{\psi_j|\hat \rho|\psi_k}=p^*_{kj}$ describe (with $j=k$) the populations or probabilities  of the state being in $\ket{\psi_j}$, or (with $j\neq k$) quantify the coherences in the system. Namely, a system whose state is $|\psi\rangle=(\ket{\uparrow}+\ket{\downarrow})/\sqrt{2}$ can be also represented as
\begin{equation}
    \hat \rho= \tfrac12\left( \ket\uparrow\bra\uparrow+\ket\uparrow\bra\downarrow+\ket\downarrow\bra\uparrow+\ket\downarrow\bra\downarrow   \right),\ \text{or equivalently}\      \rho=\frac12
    \begin{pmatrix}
    1&1\\
    1&1
    \end{pmatrix}
\end{equation}
represented on the $\{\ket\uparrow,\ket\downarrow\}$ basis.
Again, the diagonal terms are the populations of $\ket\uparrow$ and $\ket\downarrow$, while the off-diagonal ones represent the coherences. The action of decoherence models is to reduce the coherences and to lead in the asymptotic limit to 
\begin{equation}\label{7}
    \rho\longrightarrow\lim_{t\to\infty}\rho_t=\frac12
    \begin{pmatrix}
    1&0\\
    0&1
    \end{pmatrix}.
\end{equation}
The diagonalized state in Eq.\eqref{7} can represent an ensemble of states where half the time the state is $|\uparrow\rangle$ and half the time the state is $|\downarrow\rangle$: this would be perfectly compatible with what is observed after a measurement of the spin in the $\set{\ket\uparrow,\ket\downarrow}$ basis. However, the same statistical operator can represent an ensemble where half the time the state is $(|\uparrow\rangle+|\downarrow\rangle)/\sqrt{2}$ and half the time the state is 
 $(|\uparrow\rangle-|\downarrow\rangle)/\sqrt{2}$. This implies that just having a diagonalized density matrix does {\it not} guarantee that the state is collapsed into the states $|\uparrow\rangle$ or $|\downarrow\rangle$. This is the main reason why decoherence does not solve the measurement problem.

Over the years, different solutions  to the measurement problem have been suggested: Bohmian mechanics, many world interpretation, Quantum Darwinism, etc. (see \cite{carlesso2022present} and references therein). 
In this context, collapse models offer a simple way out: since the standard formulation is ambiguous in telling when postulate 1 or 2 should be applied, a unified dynamics including both dynamical postulates is employed. One modifies the Schr\"odinger equation by adding terms that mimic the collapse. Quite interestingly, due to such modifications, collapse models provide different predictions  from those of standard quantum theory. Therefore, they can be tested through experiments, as we will discuss in detail in section \ref{sec_exp}.

\section{Mathematical description of collapse models}

Spontaneous collapse models solve the measurement problem by modifying the evolution for the wavefunction, or state vector, in order to include the collapse as part of the dynamical evolution. Before introducing specific models,  we discuss the general features that collapse dynamics must fulfill. In particular, we  show why the modifications describing  the collapse must be both {non-linear} and {stochastic} \cite{bassi2003dynamical}.
%%%%%%%%

\subsection{Linear vs non-Linear modifications}
We start by considering linear modifications of the Schr\"odinger equation. 
A way to introduce a collapse-like dynamics would be to add a non-Hermitian term to the RHS of the Schr\"odinger equation:
\begin{equation}\label{non-her}
    i\hbar\frac{\D}{\D t}\ket{\psi_{t}}=\left(\hat H+i\hat D\right)\ket{\psi_{t}}
\end{equation}
where $\hat H$ is the standard Hamiltonian, while $\hat D$ is a self-adjoint operator, possibly responsible for the collapse. Indeed, such an equation would lead to complex energy eigenvalues, which in turn would lead to a decay of certain components of a superposition, namely a collapse.
However, Eq.~\eqref{non-her} does not preserve the norm of the state vector. Indeed, it is easy to show that:
\begin{equation}
    \frac{\D}{\D t}\langle\psi_{t}|\psi_{t}\rangle=\frac{2}{\hbar}\langle\psi_{t}|\hat D|\psi_{t}\rangle,
\end{equation}
and hence the probabilistic interpretation of the state is lost. To solve this, one can 
define
a normalized state via $|\phi_{t}\rangle=|\psi_{t}\rangle/\sqrt{\braket{\psi_t|\psi_t}}$. However, the corresponding dynamical equation becomes
\begin{equation}\label{nonlinD}
    i\hbar\frac{\D}{\D t}|\phi_{t}\rangle=\left[\hat H+i\left(\hat D-\langle\phi_{t}|\hat D|\phi_{t}\rangle\right)\right]|\phi_{t}\rangle,
\end{equation}
which is non-linear in the state $\ket{\phi_t}$. In order to keep the usual probabilistic interpretation of the state (i.e.~the Born rule) valid, it is necessary to work with normalized states. Then,  Eq.~\eqref{nonlinD}, which is non-linear, has  the right structure. However, as we will see below, non-linear and deterministic modifications of the Schr\"odinger equation, such as those in Eq.~\eqref{nonlinD}, allow for superluminal signalling and should thus be not considered as acceptable modifications. 

Another possible linear modification can be obtained by adding to the Hamiltonian of the system a random potential that implements a non-deterministic dynamics. The corresponding  equation for the state vector reads 
\begin{equation}\label{eq.lin1}
    i\hbar\frac{\D}{\D t}|\psi_{t}\rangle=\left(\hat H-\hbar\sqrt{\alpha}\hat A w_{t}\right)|\psi_{t}\rangle,
\end{equation}
where $w_t$ {is a stochastic process}. However, Eq.~\eqref{eq.lin1} is a standard Schr\"odinger equation, and thus it cannot lead to collapse. To see this explicitly, let us neglect the Hamiltonian evolution and study the evolution of the state vector for each realization of the noise. Namely, by taking $\psi_t(a)=\langle a|\psi_t\rangle$, where $\ket a$ is an eigenstate of $\hat A$ with associate eigenvalue $a$, we have that its evolution is given by
\begin{equation}
    \psi_{t}(a)=\psi_{0}(a)e^{-i\sqrt{\alpha}aW_{t}},
\end{equation}
where $W_{t}=\int_{0}^{t}\D s\,w_{s}$. In particular, the populations $P_t(a)=|\psi_t(a)|^2=|\psi_0(a)|^2$ are conserved, which implies that Eq.~\eqref{eq.lin1} does not describe, as expected, a wavefunction collapse.

We conclude by noticing that Eq.~\eqref{eq.lin1}, despite not being a collapse equation, leads to decoherence effects. For example, when $w_t$ is  white noise (meaning that $\mathbb E[w_t]=0$ and $\mathbb E[w_tw_s]=\delta(t-s)$, where ``$\mathbb E$" is the average over a statistical ensemble of realizations of the noise), the corresponding dynamics for the statistical operator $\hat \rho$ is of the Lindblad form: 
\begin{equation}\label{Lin1}
    \frac{\D\hat \rho(t)}{\D t}=-\frac{i}{\hbar}\left[\hat H,\hat \rho(t)\right]-\frac{\alpha}{2}\left[\hat A,\left[\hat A,\hat\rho(t)\right]\right].
\end{equation}
The last term of such a master equation leads to decoherence in the basis of the eigenstates of $\hat A$: coherences go to zero due to the different phases that they acquire at each realization of the noise.

\subsection{Non-linear and deterministic modifications}

We understood from the previous discussion that non-linear modifications are required. By employing a very elegant argument first introduced by Gisin \cite{gisin1989stochastic}, we now show that such modifications must also be stochastic as long as  no faster than light signalling is required. 
The argument goes as follows: any non-linear but deterministic modification of the Schr\"odinger equation results in a non-linear equation for the statistical operator $\hat \rho$. This leads to a dynamics that can be used to perform faster than light signalling. Thus, to avoid such a signalling, one is not allowed to have non-linear deterministic modifications of the Schr\"odinger equation.

To see this, let us focus on a thought experiment where Alice and Bob share the following singlet spin state at arbitrary distance
\begin{equation}
|\psi\rangle=\frac{|\uparrow^\text{\tiny A}\rangle|\downarrow^\text{\tiny B}\rangle-|\downarrow^\text{\tiny A}\rangle|\uparrow^\text{\tiny B}\rangle}{\sqrt{2}},
\end{equation}
where $\ket{\updownarrow^\text{\tiny A}}$  and $\ket{\updownarrow^\text{\tiny B}}$ indicate respectively the spin states of Alice and Bob alone. The state $\ket{\psi}$ is  entangled, which means that the outcomes of a spin measurement made by Alice and Bob are correlated. For example, 
if Alice measures the spin along $z$ and finds that is up, corresponding of collapsing the state in $\ket{\uparrow^\text{\tiny A}}$, then Bob's state will be with 100\% probability $\ket{\downarrow^\text{\tiny B}}$ and he will find the spin down. Now, suppose Alice performs a spin measurement before Bob does. In the first run of the experiment she measures the spin along the $z$ axis, while the second run along the $x$ axis (each run involves several measurements). Then, the statistical ensemble describing Bob's state right after the $z$ measurement is represented by the statistical operator
\begin{equation}\label{Bob z}
\hat\rho_\text{\tiny B}^{z}=\frac{1}{2}|\uparrow^\text{\tiny B}\rangle\langle\uparrow^\text{\tiny B}|+\frac{1}{2}|\downarrow^\text{\tiny B}\rangle\langle\downarrow^\text{\tiny B}|,
\end{equation}
while, after the $x$ measurement, the statistical operator describing Bob's ensemble is
\begin{equation}\label{Bob x}
\hat\rho_\text{\tiny B}^{x}=\frac{1}{2}|+^\text{\tiny B}\rangle\langle+^\text{\tiny B}|+\frac{1}{2}|-^\text{\tiny B}\rangle\langle-^\text{\tiny B}|.
\end{equation}
where $\ket{\pm^\text{\tiny B}}=(\ket{\uparrow^\text{\tiny B}}\pm\ket{\downarrow^\text{\tiny B}})/\sqrt{2}$. By explicitly substituting the latter relations in Eq.~\eqref{Bob x}, one finds that $\hat \rho_\text{\tiny B}^{z}=\hat \rho_\text{\tiny B}^{x}$, meaning that the two ensembles, while being composed by different states, are statistically equivalent. Suppose now that Bob lets the ensembles evolve and that the two initially equivalent ensembles will evolve into different ones, i.e.~$\hat \rho_\text{\tiny B}^{z}(t)\neq\hat \rho_\text{\tiny B}^{x}(t)$. Since the states can be distinguished via suitable measurements, Bob can tell which kind of measurement Alice chose to perform. This means that Alice can encode the information she wants to convey in the direction of the spin measurement (e.g., she measures along $x$ to communicate 0 and along $z$ for 1). Since Alice and Bob can be at arbitrary distances, this would allow for faster than light signaling, in plain contrast with the rules set by special relativity. 

To avoid this, the dynamical map $\mathcal M_t$ describing the evolution of the statistical operator of Bob must map uniquely any state $\hat\rho_0$ to a state $\hat\rho(t)=\mathcal M_t[\hat\rho_0]$, independently of which is the ensemble underlying $\hat\rho_0$. This, as we will see now, implies that the map $\mathcal M_t$ must be linear. 

To see why $\mathcal M_t$ must be linear, we start by noting that, as a consequence of the statistical nature of the ensemble, the states  in Eq.~\eqref{Bob z} and \eqref{Bob x} must evolve as 
\begin{equation}\label{rozt}
    \mathcal{M}_{t}[\hat{\rho}_{\text{\tiny B}}^{z}]=\frac{1}{2}|\uparrow_{t}^{\text{\tiny B}}\rangle\langle\uparrow_{t}^{\text{\tiny B}}|+\frac{1}{2}|\downarrow_{t}^{\text{\tiny B}}\rangle\langle\downarrow_{t}^{\text{\tiny B}}|,
\end{equation}
and 
\begin{equation}\label{roxt}
   \mathcal{M}_{t}[\hat{\rho}_{\text{\tiny B}}^{x}]=\frac{1}{2}|+_{t}^{\text{\tiny B}}\rangle\langle+_{t}^{\text{\tiny B}}|+\frac{1}{2}|-_{t}^{\text{\tiny B}}\rangle\langle-_{t}^{\text{\tiny B}}|,
\end{equation}
where $|\updownarrow_{t}^{\text{\tiny B}}\rangle$ and $|\pm_{t}^{\text{\tiny B}}\rangle$ are the evolved states  of the pure states $|\updownarrow^{\text{\tiny B}}\rangle$ and $|\pm^{\text{\tiny B}}\rangle$ (the evolution of which can be also non-linear). 

We now consider a third ensemble, where the experiment described above is repeated but Alice performs measurements along $z$ with a probability $p$ and along $x$ with probability $1-p$. The statistical ensemble is 
\begin{align}\label{rop}
    \hat{\rho}_{\text{\tiny B}}^{p}&=\frac{p}{2}|\uparrow^{\text{\tiny B}}\rangle\langle\uparrow^{\text{\tiny B}}|+\frac{p}{2}|\downarrow^{\text{\tiny B}}\rangle\langle\downarrow^{\text{\tiny B}}|+\frac{1-p}{2}|+^{\text{\tiny B}}\rangle\langle+^{\text{\tiny B}}|+\frac{1-p}{2}|-^{\text{\tiny B}}\rangle\langle-^{\text{\tiny B}}|=\nonumber\\
    &=p\;\hat{\rho}_{\text{\tiny B}}^{z}+(1-p)\hat{\rho}_{\text{\tiny B}}^{x},
\end{align}
which is again equivalent to those in Eqs.~\eqref{Bob z} and \eqref{Bob x}, i.e. $\hat{\rho}_{\text{\tiny B}}^{p}=\hat{\rho}_{\text{\tiny B}}^{x}=\hat{\rho}_{\text{\tiny B}}^{z}$. 
Also in this case, due to the statistical nature of the ensemble, the state in Eq.~\eqref{rop} must evolve as:
\begin{align}
\mathcal{M}_{t}[\hat{\rho}_{\text{\tiny B}}^{p}]&=\frac{p}{2}|\uparrow_{t}^{\text{\tiny B}}\rangle\langle\uparrow_{t}^{\text{\tiny B}}|+\frac{p}{2}|\downarrow_{t}^{\text{\tiny B}}\rangle\langle\downarrow_{t}^{\text{\tiny B}}|+\frac{1-p}{2}|+_{t}^{\text{\tiny B}}\rangle\langle+_{t}^{\text{\tiny B}}|+\frac{1-p}{2}|-_{t}^{\text{\tiny B}}\rangle\langle-_{t}^{\text{\tiny B}}|\nonumber\\
&=p\mathcal{M}_{t}[\hat{\rho}_{\text{\tiny B}}^{z}]+(1-p)\mathcal{M}_{t}[\hat{\rho}_{\text{\tiny B}}^{x}],
\end{align}
which, by comparing with the second line of Eq.~\eqref{rop}, implies the map $\mathcal M_t$ is linear. 

In summary, to guarantee that equivalent ensembles are mapped to equivalent ones, thus making impossible for Bob to distinguish which measurements Alice performed and hence not allowing for  faster than light signalling, the evolution map $\mathcal M_t$ must be linear.

\subsection{The Ghirardi-Rimini-Weber  model}
\begin{figure}[ht!]
    \centering
    \includegraphics[width=1\linewidth]{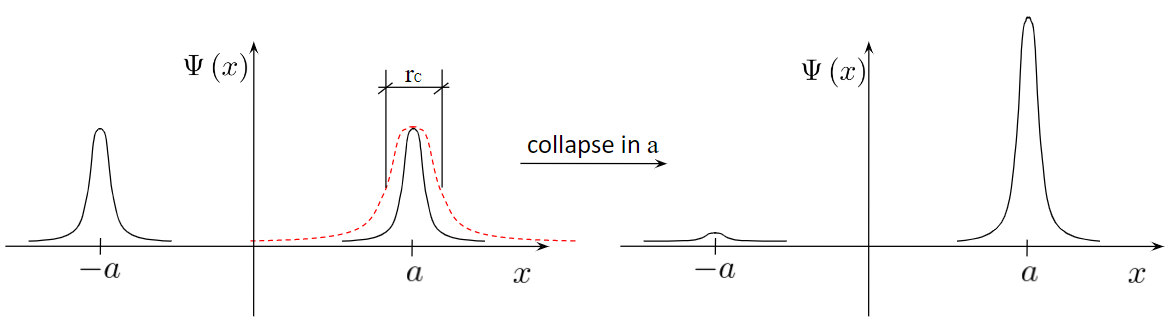}
    \caption{A superposition of two wavepackets, centered respectively around ${\bf a}$ and $-{\bf a}$ (black continuous lines), is subject to a GRW spontaneous localization around the point ${\bf a}$ (red dashed line), leading to a collapse of the superpostion around ${\bf a}$. 
       \label{fig:colGRW}}
\end{figure}

The first established 
 collapse model appearing in the literature was the Ghirardi-Rimini-Weber (GRW) model \cite{ghirardi1986unified}. In this model, the state vector $|\psi\rangle$ follows a piece-wise evolution: the state evolves according to the standard Schr\"odinger evolution until the latter is interrupted by sudden collapse which localises the system in position. 
More precisely, the postulates of the model are the following:
\begin{enumerate}
    \item The wavefunction is subject to spontaneous localizations in space that take place 
at random times. Their time distribution is described by a Poisson statistic with rate $\lambda$. 
In the original GRW paper, one has different parameters $\lambda_\kappa$, one for each type of particle.
    \item The localization, or collapse, of the $i$-th particle around the position ${\bf a}$ is described by
    \begin{equation}\label{GRW.loc}
|\psi\rangle\longrightarrow\frac{\hat L_{{\bf a}}^{(i)}|\psi\rangle}{||\hat L_{{\bf a}}^{(i)}|\psi\rangle||},
    \end{equation}
    where $||\ket\phi||=\sqrt{\braket{\phi|\phi}}$ and we defined the collapse operator
    \begin{equation}\label{L}
\hat L_{{\bf a}}^{(i)}=\frac{1}{(\pi \rC^{2})^{3/4}}e^{-{(\hat{\q}_{i}-{\bf a})^{2}}/{2\rC^{2}}},
    \end{equation}
    where $\hat \q_i$ is the position operator of the $i$-th particle and the parameter $\rC$ describes the length-scale above which the collapse is more effective.
    \item The probability of having a localization around the point ${\bf a}$ is given by 
    \begin{equation}\label{GRW.prob}
P({\bf a})=||\hat L_{{\bf a}}^{(i)}|\psi\rangle||^{2}.
    \end{equation}
    We emphasize that the choice made for the collapse operator $\hat L_{{\bf a}}^{(i)}$ in Eq.~\eqref{L} guarantees that the probabilities sum to one, i.e.~$\int \D{\bf a}\,P({\bf a})=1$.
    \item Between two consecutive localizations, the state evolves according to the standard Schr\"odinger equation \eqref{sch}. 
\end{enumerate}
We dwell now -- through some examples -- 
in how the collapse works in the GRW model. 
First, we show that the collapse localises the wavefunction in space. Let us suppose that the state is in a  superposition of two wavepackets respectively localised at the points $-{\bf a}$ and ${\bf a}$. This can be represented as
\begin{equation}
\psi({\bf x})=\frac{1}{\mathcal{N}}\left(e^{-\frac{({\bf x}-{\bf a})^{2}}{2\Delta^{2}}}+e^{-\frac{({\bf x}+{\bf a})^{2}}{2\Delta^{2}}}\right),
\end{equation}
where $\mathcal{N}$ is the appropriate normalization factor. For the sake of simplicity, we suppose that $\Delta\ll \rC\ll |{\bf a}|$. Let us suppose that a localization takes place at the position ${\bf a}$. Then, following the postulate number 2, 
 the part of $\psi({\bf x})$ described by the Gaussian centered in ${\bf a}$
 will be (to a very good approximation) unaffected by the collapse. Conversely, the other part, being centred in $-{\bf a}$, will be highly suppressed. We notice that due to the postulate number 3, there is a very high probability that the localizations take place around positions $-{\bf a}$ and ${\bf a}$, while the chances of having localization in other positions is quite low. Figure \ref{fig:colGRW} describes visually the collapse for this example.

To be a good collapse model,  the collapse effects need to be negligible for microscopic systems, like atoms or molecules, for which we know that the Schr\"odinger equation works extremely well. Conversely, when more and more atoms bind together to form  macroscopic systems, an amplification mechanism, built in the collapse dynamics, must guarantee a strong and fast collapse, which dominates the dynamics
so to suppress macroscopic superpositions. To better understand how such an amplification mechanism works in the GRW model, let us consider a  cat state, which corresponds to the superposition of the centers of mass of $N$ particles in two distinct positions and reads
\begin{equation}\label{cat}
|\psi\rangle=\frac{|\phi_{-{\bf a}}\phi_{-{\bf a}}\dots\phi_{-{\bf a}}\rangle+|\phi_{{\bf a}}\phi_{{\bf a}}\dots\phi_{{\bf a}}\rangle}{\sqrt{2}},
\end{equation}
where $|\phi_{{\bf a}}\phi_{{\bf a}}\dots\phi_{{\bf a}}\rangle=|\phi_{{\bf a}}\rangle\otimes|\phi_{{\bf a}}\rangle\otimes\dots\otimes|\phi_{{\bf a}}\rangle$ with $|\phi_{{\bf a}}\rangle$ describing a wavepacket of a single particle, being well localised around the position ${\bf a}$. 
We immediately see that if even just one of the atoms localises due to the collapse dynamics, due to the form of the state $\ket\psi$, all the others will also get localised. Therefore, given a system of $N$ particles, if each one of them localises at the rate $\lambda$, the the state in Eq.~\eqref{cat} collapses with a rate  $\Lambda=N\lambda$. Namely the larger  the system (and thus $N$) the stronger  the collapse-induced localization.
The same amplification mechanism applies to solid objects. In particular, it can be shown that the center of mass of a rigid body composed of $N$ particles localises with the amplified rate $\Lambda$. 
We point out that the entanglement residing  in the state in Eq.~\eqref{cat} plays a crucial role for the amplification mechanism to be effective. To prove this, we can consider the  opposite case, where we have $N$ particles, all in a superposition of distinct positions, but in a separable state, i.e. 
\begin{equation}\label{nocat}
|\psi'\rangle=\left(\frac{|\phi_{-{\bf a}}\rangle+|\phi_{{\bf a}}\rangle}{\sqrt{2}}\right)\otimes\left(\frac{|\phi_{-{\bf a}}\rangle+|\phi_{{\bf a}}\rangle}{\sqrt{2}}\right)\otimes\dots\otimes\left(\frac{|\phi_{-{\bf a}}\rangle+|\phi_{{\bf a}}\rangle}{\sqrt{2}}\right).
\end{equation}
One sees immediately that the localization of one among the particles, 
 in either $|\phi_{{\bf a}}\rangle$ or $|\phi_{-{\bf a}}\rangle$, does not affect the state of the others. This is precisely due to the fact that 
 the state in Eq.~\eqref{nocat} is fully separable.

To conclude our discussion on the GRW model, we derive the corresponding master equation for the statistical operator $\hat \rho$. In terms of $\hat \rho=\ket\psi\bra\psi$, the action of the localization process in Eq.~\eqref{GRW.loc} for the $i$-th particle  is given by map 
\begin{equation}
    \mathcal{T}_i\left[|\psi\rangle\langle\psi|\right]=\int \D{\bf a}\,P({\bf a})\frac{\hat L_{{\bf a}}^{(i)}|\psi\rangle\langle\psi|\hat L_{{\bf a}}^{(i)\dagger}}{||\hat L_{{\bf a}}^{(i)}|\psi\rangle||^{2}}=\int \D{\bf a}\,\hat L_{{\bf a}}^{(i)}|\psi\rangle\langle\psi|\hat L_{{\bf a}}^{(i)\dagger},
\end{equation}
where Eq.~\eqref{GRW.prob} was employed. Since such a map is linear, it can be extended to any state $\hat \rho$. 
Now, given the state $\hat\rho(t)$ at time $t$, we compute that at time $t+\D t$.
Since the localizations take place following a Poisson statistic in the time, there is a probability of  localization of $\lambda \D t$  per particle. Correspondingly, one has a probability of $(1-N\lambda \D t)$ that there are no localizations and the system evolves according to the standard Schr\"odinger evolution. This means the state at time $t+\D t$ is given by
\begin{equation}
 \hat \rho(t+\D t)=(1-N\lambda \D t)\left(\hat \rho(t)-\frac{i}{\hbar}\left[\hat H,\hat\rho(t)\right]\D t\right)+\lambda \D t\sum_i\mathcal{T}_i\left[\hat \rho(t)\right],
\end{equation}
from which one easily derives the GRW master equation
\begin{equation}\label{GRW_ME}
    \frac{\D\hat\rho(t)}{\D t}=-\frac{i}{\hbar}\left[\hat H,\hat\rho(t)\right]+\lambda\sum_i\left(\int \D{\bf a}\,\hat L_{{\bf a}}^{(i)}\hat\rho(t)\hat L_{{\bf a}}^{(i)\dagger}-\hat\rho(t)\right),
\end{equation}
where terms of order $\D t^2$ are neglected.

\subsection{The Continuous Spontaneous Localisaton model}

The GRW model presents two limitations. The first one, more aesthetic, is that the dynamics of the state vector is not defined by a single equation but by a set of postulates. The second limitation, which is more substantial, is that the collapse does not preserve the symmetry of the state when considering systems of identical particles. 

Both these limits were taken into account  with the introduction of the Continuous Spontaneous {Localization} (CSL) model, which was formulated by Ghirardi, Rimini and Pearle in 1990 \cite{ghirardi1990markov}. 
In the CSL model the collapse is continuous in time, described by a stochastic differential equation of the same form of those used in the theory of continuous measurements. Before introducing the CSL model, we introduce the general properties of continuous collapse equations. The collapse equation for the state vector reads  
\begin{equation}\label{ito}
\D |\psi_{t}\rangle=\left[-\frac{i}{\hbar}\hat H \D t+\sqrt{\gamma}\sum_{i=1}^{N}(\hat A_{i}-\langle \hat A_{i}\rangle_{t})\D W_{i,t}-\frac{\gamma}{2}\sum_{i=1}^{N}(\hat A_{i}-\langle \hat A_{i}\rangle_{t})^{2}\D t\right]|\psi_{t}\rangle,
\end{equation}  
where $\gamma$ sets the strength of the collapse mechanism and $\D W_{i,t}$ is a set of $N$ independent Wiener increments, with zero average and correlations $\mathbb{E}[\D W_{i,t}\D W_{j,t}]=\delta_{i,j}\D t$, with $\mathbb{E}[\dots]$ denoting the average over the Wiener processes. $\hat A_{i}$ are a set of commuting operators and $\langle \hat A_{i}\rangle_{t}=\langle \psi_t| \hat A_{i}|\psi_t\rangle$ denotes their quantum averages on the state $|\psi_t\rangle$.
The dynamics in Eq.~\eqref{ito} has two important properties:
\begin{enumerate}
    \item The second and the third terms induce a dynamics that collapses the state into an eigenstate of one of the operators $\hat A_i$.
    \item The master equation associated with the state vector dynamics in Eq.~\eqref{ito} is 
    \begin{equation}\label{CSL_ME}
    \frac{\D  \hat\rho(t)}{\D t}=-\frac{i}{\hbar}\left[\hat H,\hat \rho(t)\right]-\gamma\sum_{i=1}^{N}\left[\hat A_{i},\left[\hat A_{i},\hat \rho(t)\right]\right].
    \end{equation}
\end{enumerate}
A scratch of the proof of the first point is the following \cite{adler2007collapse}. 
Consider Eq.~\eqref{ito} in a regime where 
the Hamiltonian contribution to the evolution can be neglected. 
For any operator $\hat B$ that commutes with all $\hat A_i$, one can show that the variance $ V_{B}(t)=\langle \hat B^{2}(t)\rangle-\langle \hat B(t)\rangle^{2}$ is described by 
\begin{equation}\label{var}
\mathbb{E}\left[ V_{B}(t)\right]= V_{B}(0)-4\gamma\sum_{i=1}^{N}\int_{0}^{t} \D s\,\mathbb{E}\left[ C_{
B,A_{i}}(s)\right],
\end{equation}
where $ C_{
B,A_{i}}(t)=\langle (\hat B-\langle \hat B\rangle_{t})(\hat A_{i}-\langle \hat A_{i}\rangle_{t})\rangle _{t}$ is the correlation between the operators $\hat B$ and $\hat A_i$.
In the case that $\hat B=\hat A_j$ is one of the collapse operators, the $j$-th term in the sum in the right hand side of Eq.~\eqref{var} is of the form $-4\gamma\int_{0}^{t}\D s\,\mathbb{E}\left[ V_{A_{j}}(s)\right]$. The latter
converges in the limit $t\to \infty$ only if 
$\mathbb{E}[ V_{A_{j}}(t)]\to 0$. However, since 
$\mathbb{E}[ V_{A_{j}}(t)]$ is the average, over different realizations of the noise, of positive quantities (indeed, $V_{A_{j}}(t)$ are variances computed on the single statevector), then $V_{A_{j}}(t)$ needs to vanish for each realization.
This implies that the statevector  is an eigenstate of $\hat A_j$. This  applies for any $j$, and proves that the non-Hamiltonian part of the dynamics leads to collapse the state in a common eigenstates of all the operators $\hat A_j$.

To show the second point, we study the infinitesimal evolution
of the statistical operator $\hat\rho(t)=\mathbb{E}\left[|\psi_{t}\rangle\langle\psi_{t}|\right]$ associated with Eq.~\eqref{ito}. The latter is given by
\begin{align}\label{der_Lin}
    \D \hat \rho(t)&=\D \mathbb{E}\left[|\psi_{t}\rangle\langle\psi_{t}|\right]=\mathbb{E}\left[|\D \psi_{t}\rangle\langle\psi_{t}|+|\psi_{t}\rangle\langle \D \psi_{t}|+|\D \psi_{t}\rangle\langle \D \psi_{t}|\right]=\\
    &=-\frac{i}{\hbar}\hat H \hat \rho(t)\D t-\frac{\gamma}{2}\sum_{i=1}^N \hat A_{i}^2\hat \rho(t)\D t+
    %\nonumber
    %\\
    \frac{i}{\hbar}\hat \rho(t)\hat H\D t-\frac{\gamma}{2}\sum_{i=1}^N\hat \rho(t)\hat A_{i}^2\D t+\nonumber
    \\
&+\gamma\sum_{i,j=1}^N\delta_{i,j}\hat A_{i}\hat \rho(t)\hat A_{j}\D t,\nonumber
\end{align}
where we used the rules of It\^o calculus,  in particular
\begin{equation}
\mathbb{E}\left[\hat A_{i}\D W_{i,t}|\psi_{t}\rangle\langle\psi_{t}|\right]=0,
\end{equation}
and 
\begin{equation}
\mathbb{E}\left[\hat A_{i}\D W_{i,t}|\psi_{t}\rangle\langle\psi_{t}|\D W_{j,t}\hat A_{j}\right]=\hat A_i\hat \rho(t)\hat A_j\delta_{i,j}\D t.
\end{equation}
The master equation \eqref{CSL_ME} follows immediately.

Finally, we note an important feature of stochastic dynamics. The master equation \eqref{CSL_ME} can be derived also from a stochastic Schr\"odinger equation of the form \cite{bassi2003dynamical}:
\begin{equation}\label{ss}
i\hbar\frac{\D |\psi_{t}\rangle}{\D t}=\left[\hat H-\hbar\sqrt{\gamma}\sum_{i=1}^{N}\hat A_{i}w_{i,t}\right]|\psi_{t}\rangle,
\end{equation}  
where $w_{i,t}=\D W_{i,t}/\D t$ are white noises with zero average and correlation $\mathbb{E}[w_{i,t}w_{j,t'}]=\delta_{ij}\delta(t-t')$. Note that Eq.~\eqref{ss} is written in the Stratonovich form, in contrast to Eq.~\eqref{ito} which is written in the It\^o form. 
While Eq.~\eqref{ss}, which is a standard Schr\"odinger equation, cannot be used to describe the wavefunction collapse, it is extremely useful when one computes predictions to be compared with experiments: indeed, since Eq.~\eqref{ito} and Eq.~\eqref{ss} lead to the same master equation \eqref{CSL_ME}, they also provide the same  expectation values for any operator.

We now proceed by specializing Eq.~\eqref{ito} to the case of the CSL model.
In such a case, the discrete index $i$ becomes a continuous label $\x$ corresponding positions in space, namely
\begin{equation}\label{CSL2nd}
\begin{aligned}
\D|\psi_{t}\rangle=\left[-\frac{i}{\hbar}\hat H\D t+\frac{\sqrt{\gamma}}{m_{0}}\int \D \x\,(\hat M(\x)-\langle \hat M(\x)\rangle_{t})\D W_{t}(\x)\right.\\
\left.-\frac{\gamma}{2m_{0}^{2}}\int \D\x\,\left(\hat M(\x)-\langle \hat M(\x)\rangle_{t}\right)^{2}\D t\right]|\psi_{t}\rangle,
\end{aligned}
\end{equation}
where the collapse operators are 
\begin{equation}\label{M2nd}
\hat M(\x)=\sum_{j,s}m_{j}\int \D \y\,g(\x-\y)\hat\psi_{j}^{\dagger}(\y,s)\hat \psi_{j}(\y,s),
\end{equation}
where $m_j$ is the mass of the particles of the $j$-th kind,
\begin{equation}
    g(\x-\y)=\frac{1}{(\sqrt{2\pi}\rC)^{3}}e^{-\frac{(\x-\y)^{2}}{2\rC^{2}}},
\end{equation}
and $\hat\psi_{j}^\dagger(\y,s)$ and $\hat\psi_{j}(\y,s)$ are, respectively, the creation and annihilation operators of a particle of the kind $j$ with spin $s$. The Wiener increments associated with different points in space are independent, so $\mathbb{E}[\D W_{t}(\x)]=0$ and $\mathbb{E}[\D W_{t}(\x)\D W_{t}(\y)]=\delta(\x-\y)\D t$.
In the typical case of a system where the number of particles is fixed and  the symmetry of the wavefunction and the spins of particles do not play any crucial role, one can consider a version of the CSL model in the first quantization formalism. In such a case, the operators in Eq.~\eqref{M2nd} become
\begin{equation}\label{M1st}
\hat M(\x)=\sum_{i=1}^{N}m_{i}\,g(\x-\hat{\q}_{i}),
\end{equation}
where now the sum runs over the $N$ particles composing the system. 

Sometimes it is also useful to rewrite the CSL dynamical equation in the following alternative form
\begin{align}\label{CSL_gau}
&\D |\psi_{t}\rangle=\left[-\frac{i}{\hbar}\hat H\D t+\frac{\sqrt{\lambda}}{m_{0}}\int \D \x\,\left(\hat\mu(\x)-\langle\hat\mu(\x)\rangle_{t}\right)\D \overline{W}_{t}(\x)\right.\\
&\left.-\frac{\lambda}{2m_{0}^{2}}\int \D \x\int \D \y\,G(\x-\y)(\hat\mu(\x)-\langle\hat\mu(\x)\rangle_{t})(\hat\mu(\y)-\langle\hat\mu(\y)\rangle_{t})\D t\right]|\psi_{t}\rangle,\nonumber
\end{align}
where $\hat\mu(\x)=\sum_{i=1}^{N}m_{i}\delta(\x-\hat{\q}_{i})$ is the mass density of the system and the Wiener increments $\D \overline{W}_{t}$ have a Gaussian correlation reading
\begin{equation}\label{GCSL}
\mathbb{E}[\D \overline{W}_{t}(\x)\D \overline{W}_{t}(\y)]= G(\x-\y)\D t,
\end{equation}
where
$G(\x-\y)=e^{-{(\x-\y)^{2}}/{4\rC^{2}}}
$.

The master equation of the CSL model, with respect to the definition of the collapse operators given in Eq.~\eqref{M1st}, reads: 
\begin{equation}\label{CSLME}
\frac{\D \hat\rho(t)}{\D t}=-\frac{i}{\hbar}\left[\hat H,\hat\rho(t)\right]-\frac{\lambda}{2\pi^{3/2}m_{0}^{2}\rC^{3}}\int \D\x\,\left[\hat M(\x),\left[\hat M(\x),\hat\rho(t)\right]\right].
\end{equation}
It can be shown that, in the case of a single particle,  this master equation is identical to the GRW one displayed in Eq.~\eqref{GRW_ME}. 
Conversely, when many particles are considered, the two models show differences. An example is given by a different amplification mechanism: while in the GRW model the center of mass of a rigid body composed of $N$ particles always collapses  with an amplified rate $\Lambda_\text{\tiny GRW}=N\lambda$, in the CSL model the amplified rate depends on the distance between the particles composing the system with respect to $\rC$. Two limiting cases are relevant: one can show that when $\rC$ is much smaller than the distance between the particles, one gets $\Lambda_\text{\tiny CSL}\simeq N\lambda$, as for the GRW model; on the other hand, when $\rC$ is much larger than the size of the system, one gets a quadratic amplification in the number of particles $\Lambda_\text{\tiny CSL}\simeq N^2\lambda$, which scales differently compared to the GRW one. 
Finally, we note that in several practical situations, the system under study can be  approximated as a rigid body,  with the motion of its centre of mass  on length-scales smaller than $\rC$. In such a regime, one can approximate Eq.~\eqref{CSLME} to
\begin{equation}\label{lin}
\frac{\D \hat{\rho}(t)}{\D t}=-\frac{i}{\hbar}\left[\hat{H},\hat{\rho}(t)\right]-\sum_{i,j}\eta_{ij}\left[\hat{q}_{i},\left[\hat{q}_{j},\hat{\rho}(t)\right]\right],
\end{equation}
where $\hat \q$ is the center of mass position operator and 
\begin{equation}\label{eta}
\eta_{ij}=\frac{\lambda \rC^{3}}{2\pi^{3/2}m_{0}^{2}}\int \D \k\,|\tilde{\mu}(\k)|^2k_{i}k_{j}e^{-k^{2}\rC^{2}},
\end{equation}
is a the diffusive coefficient with $\tilde{\mu}(\k)=\int \D\r\, e^{-i\k\cdot\r}\mu(\r)$ being the Fourier transform of the mass density $\mu(\r)$.

\subsection{The Di\'osi-Penrose model}
In the GRW and the CSL models, the spontaneous collapse is postulated to solve the measurement problem. However, there is no explanation or indication about its origin. An attempt in this direction was given, following different arguments, by Di\'osi and Penrose. 
To date, gravity is the only fundamental interaction that has not been successfully quantized. Thus, one can suppose that gravity is different: it should not be quantized and is at the origin of  wavefunction collapse \cite{donadi2022seven}.
Two interesting proposals in this direction were put forward by Di\'osi \cite{diosi1989models} and Penrose \cite{penrose1996gravity}. Despite being based on different ideas, these proposals arrive at the same formula for the time of collapse of a spatial superposition. For this reason, they are commonly known under the unique name of the Di\'osi-Penrose model.
We will introduce them separately, to show the common points and the differences. 

\paragraph{The Di\'osi model}
The Di\'osi model was introduced in 1989 \cite{diosi1989models}, and it is a collapse model where the state vector evolves according to an equation of the form~\eqref{ito}, with the collapse operators being the {mass density, and the noise} having a spatial correlation related to  Newtonian gravity. Namely, its dynamical equation reads
\begin{align}\label{diosi}
    \D |\psi(t)\rangle&=\left[-\frac{i}{\hbar}\hat{H}\D t+\sqrt{\frac{G}{\hbar}}\int \D \r\left(\hat{\mu}(\r)-\langle\hat{\mu}(\r)\rangle_{t}\right)\D W_{t}(\r)\right.\\
    &\left.-\frac{G}{2\hbar}\int \D \r\int \D \r'\,\frac{\left(\hat{\mu}(\r)-\langle\hat{\mu}(\r)\rangle_{t}\right)\left(\hat{\mu}(\r')-\langle\hat{\mu}(\r')\rangle_{t}\right)}{|\r-\r'|}\right]|\psi(t)\rangle,\nonumber
\end{align}
where $\hat{\mu}(\r)$ is the mass density operator defined after Eq.~\eqref{CSL_gau} and the Wiener processes have correlation $\mathbb{E}[\D W_{t}(\r)\D W_{t}(\r')]=1/|\r-\r'| \,\D t$.
The corresponding master equation is 
\begin{equation}\label{masterD}
    \frac{\D \hat{\rho}(t)}{\D t}=-\frac{i}{\hbar}\left[\hat{H},\hat{\rho}(t)\right]-\frac{G}{2\hbar}\int \D \r\int \D \r'\frac{1}{|\r-\r'|}\left[\hat{\mu}(\r),\left[\hat{\mu}(\r'),\hat{\rho}(t)\right]\right].
\end{equation}
To study how fast the model collapses in space, one can neglect the Hamiltonian evolution and look at the dynamics of the off-diagonal elements in the position basis, which reads
\begin{equation}
\langle\x|\hat{\rho}(t)|\y\rangle=\langle\x|\hat{\rho}(0)|\y\rangle e^{-t/\tau_\text{\tiny D}},
\end{equation}
where we introduced the time of decay $\tau_{D}$ defined as
\begin{equation}\label{tau}
\tau_\text{\tiny D}^{-1}=\frac{G}{2\hbar}\int \D \r\int \D \r'\,\frac{[\mu(\r-\x)-\mu(\r-\y)][\mu(\r'-\x)-\mu(\r'-\y)]}{|\r-\r'|},
\end{equation}
with $\mu(\r-\x)=\sum_{i=1}^{N}m_{i}\delta(\r-\x_i)$.  Equation \eqref{tau} indicates that objects with large masses collapse faster compared to small ones: to give an estimation, the time of decay for a proton is $\tau_\text{\tiny D}\simeq 10^6$\,years, while for a dust of grain (assumed to be spherical) with a  radius of $10\,\mu$m and a mass of $10^{-12}$\,kg one gets $\tau_\text{\tiny D}\simeq 10^{-8}$\,s.

\paragraph{Penrose's proposal}
In 1996, Penrose suggested the idea that there is a fundamental tension between the principles of General relativity and Quantum theory \cite{penrose1996gravity}. 
The main idea is the following. Suppose there is a mass in a spatial superposition of two different locations which we denote with ${\bf a}$ and ${\bf b}$. Each branch of the superposition bends spacetime in a different way, thus generating a superposition of two different spacetime metrics. According to the equivalence principle, which lies at the root of general relativity, it is always possible, by performing an appropriate change of coordinate, to find locally a free-falling reference frame where the effects of gravity are nullified and the spacetime metric is flat. However, if the metric is not well defined because of being in a superposition, this is no longer possible. Hence, there is a conflict between the superposition principle and the equivalence principle. Penrose proposed that nature resolves this conflict by collapsing the mass into a localized state, either to ${\bf a}$ or ${\bf b}$. The larger  the difference between the two generated spacetime metric, the faster the superposition is suppressed. 
Penrose suggested that, in a non-relativistic regime, a good measure of the difference between two spacetime metrics is given by
\begin{equation}
\Delta E_{\text{\tiny DP}}=\frac{1}{G}\int \D \r\,\left|\g_{{\bf a}}(\r)-\g_{{\bf b}}(\r)\right|^{2},
\end{equation}
where $\g_\x(\r)$ is the gravitational acceleration experience by a test mass at the point $\r$ when the mass generating the gravitational field is located at the point $\x$. By using $\g_\x(\r)=-\nabla\phi_\x(\r)$ and the Poisson equation $\nabla^{2}\phi_\x(\r)=4\pi G\mu_\x(\r)$, one gets
\begin{equation}
\Delta E_{\text{\tiny DP}}=4\pi G\int \D \r\int \D \r'\,\frac{\left[\mu_{{\bf a}}(\r)-\mu_{{\bf b}}(\r)\right]\left[\mu_{{\bf a}}(\r')-\mu_{{\bf b}}(\r')\right]}{|\r-\r'|}.
\label{eq:De Pen}
\end{equation}
Penrose then suggests that the time of collapse is given by $\tau=\hbar/\Delta E$, which is  precisely the one found by Di\'osi, up to a factor $8\pi$.

We note that Eq.~\eqref{tau} and Eq.~\eqref{eq:De Pen} are problematic for point-like particles. Indeed, the integrals diverge, leading to $\tau=0$. This would imply an instantaneous collapse for microscopic particles, which is clearly not the case according to experimental evidence. For this reason, one considers  course grained mass densities by introducing a length cut-off $R_0$, which gives a finite extension to mass densities of point-like particles \cite{diosi1989models}. In the case of protons, the natural value would be $R_0=10^{-15}$\,m, which corresponds to the value of the proton radius as measured by spectroscopy  and nuclear scattering experiments. However, it was shown that such a choice is  incompatible with observation: the model predicts an increase in kinetic energy, which is typical of all collapse models (see section \ref{non-int} for a more detailed discussion), and incompatible with observation.

Another suggestion is that the mass density should be related to $|\psi|^2$ and that the value of $R_0$ could be estimated from the size of the wavefunction of the system. However, this proposal was ruled out by an experiment studying the radiation emission from Germanium \cite{donadi2020underground} (see also section \ref{sec_exp}). 
The current practice is to consider
$R_0$ as a free parameter of the model, and try to place experimental bounds on it. 
The best bound on this parameter comes from the study of spontaneous radiation emission from Germanium and is given by $R_0\geq (4.94\pm 0.15)\times 10^{-10}$\,m \cite{arnquist2022search}.

\section{Experimental tests}\label{sec_exp}

After having introduced the theoretical framework of collapse models in the sections above, we can now focus on the experimental characterization of the collapse effects. There are two kind of experiments that one can exploit:  interferometric tests and  non-interferometric tests.

\subsection{Interferometric tests}

\begin{figure} [h]
   \begin{center}
   %\begin{tabular}{c} %% tabular useful for creating an array of images 
   \includegraphics[width=0.8\linewidth]{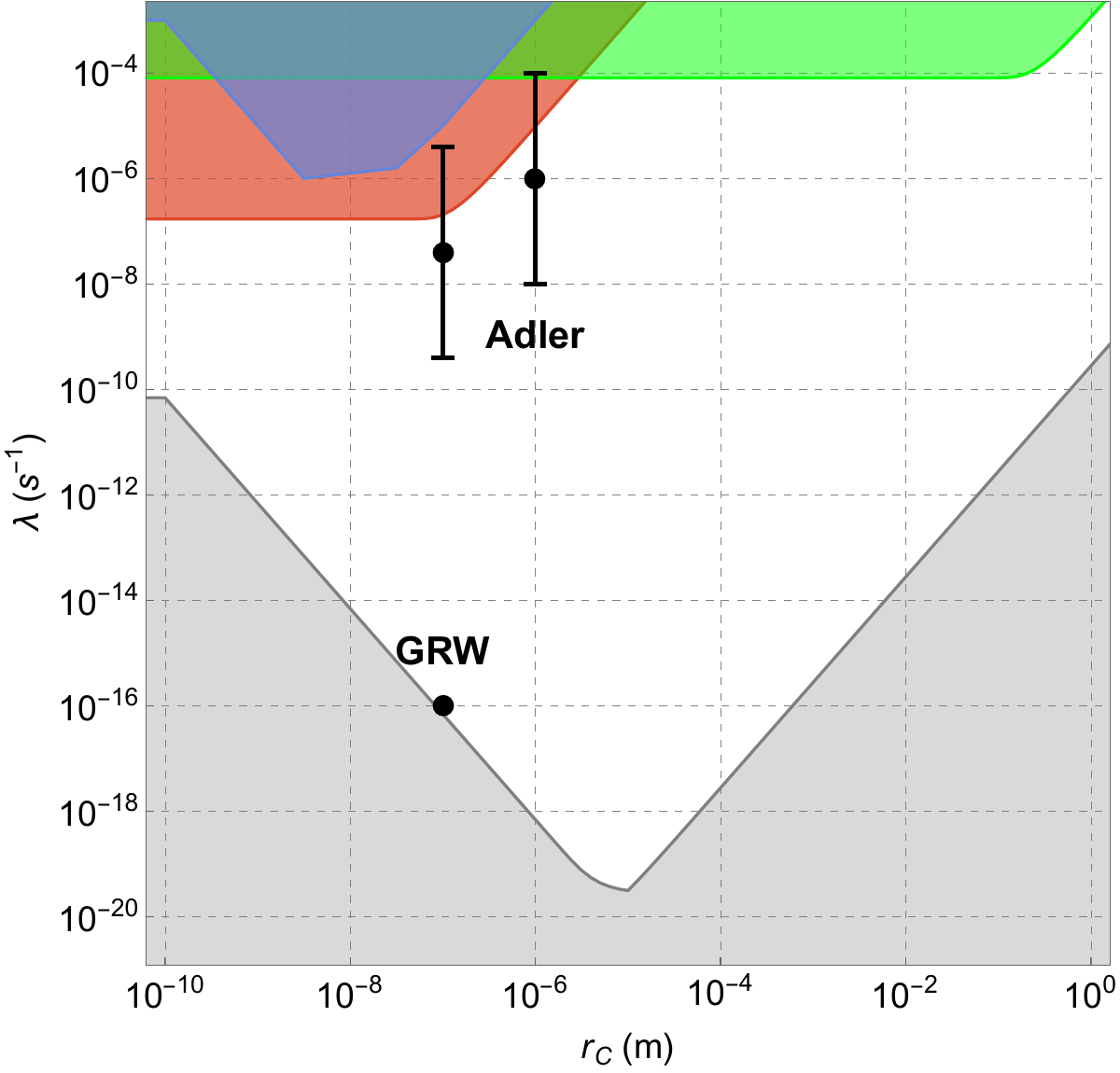}
   %\end{tabular}
   \end{center}
   \caption[example] 
%>>>> use \label inside caption to get Fig. number with \ref{}
   { \label{fig1}
Experimental upper bounds on CSL parameters $\lambda$ and $\rC$ from interferometric experiments.  The green region is excluded from {cold atom experiment}~\cite{kovachy2015quantum}. The blue and red regions are excluded by two different  molecular interferometry experiments~\cite{torovs2017colored}. The theoretical values proposed by GRW\cite{ghirardi1986unified} and the ranges proposed
by Adler\cite{adler2007lower} are shown respectively as a black dot and black dots
with bars that indicate the estimated range. Finally, the light grey
area is theoretically excluded\cite{torovs2017colored}. The white area has not been explored with interferometric experiments.
}
\end{figure}

The most natural and direct way to test collapse models is provided by interferometric tests. These directly probe the quantum superposition principle, and characterize its limits. The predictions of collapse models are to perturb, suppress and eventually destroy quantum superpositions of massive systems.
The idea here is to prepare a system in a spatial superposition (say, of two spatially distinguishable wavefunctions $\psi_1(x)$ and $\psi_2(x)$), let it freely evolve for a certain time $t_1$, re-collimate the different components of the superposition to let them interfere at time $t_2$, and finally measure the corresponding interference pattern at $\tau=t_1+t_2$. During the time $\tau$, the collapse noise acts on the superposition by perturbing it. Such an effect is reflected in the degradation of the interference pattern predicted by standard quantum  mechanics in the ideal case of a noise-free environment.
To be quantitative, if we consider a single particle of mass $m$ under free evolution for a time $\tau$, 
 the CSL model predicts a modified dynamics leading to the following form for the density matrix in the position representation \cite{torovs2017colored} 
\begin{equation}\label{eq.rho_free_CSL}
\rho(\x,\y,\tau)=\frac{1}{(2\pi\hbar)^3}\int\D\k\int\D \z \,e^{-\tfrac i\hbar \k\cdot\z}F_\text{\tiny CSL}(\k, \x-\y,\tau)\rho_\text{\tiny QM}(\x+\z,\y+\z, \tau),
\end{equation}
where $\rho_\text{\tiny QM}(\x,\y, t)$ is the density matrix as evolved under the unitary standard quantum mechanics, and
\begin{equation}\label{Finterf.CSL}
    F_\text{\tiny CSL}(\k,\q,t)=\exp\left[-\lambda\frac{m^2}{m_0^2}\left(t-\int_0^t\D s\,e^{-(\q-\k s/m)^2/4\rC^2}\right)\right],
\end{equation}
quantifies the action of the CSL noise. To account for more complex situations, for example in the presence of an external potential or when preparing/collimating the superposition, the expression in Eq.~\eqref{eq.rho_free_CSL} needs to be suitably modified. Nevertheless, such an equation already provides all the indications of the collapse action and which are the important dependencies.
To amplify the collapse effect, one can increase \textit{i)} the mass $m$ of the interfered system, \textit{ii)} the time $\tau$ for which the collapse noise can act, and \textit{iii)} the superposition distance $\Delta_x=|\braket{\psi_1|\hat x|\psi_1}-\braket{\psi_2|\hat x|\psi_2}|$. Preparing and maintaining over time spatial superposition of massive systems is challenging due to the technical requirements on isolation from environmental influences that might spoil the superposition. Currently, {the record for the most massive particle placed in a superposition of distinguishable positions is around} $10^4\,\text{amu}\simeq 10^{-23}$\,kg which was  attained with macromolecule interferometric experiments, and led to bounding CSL to $\lambda<10^{-7}\,$s$^{-1}$ at $\rC=10^{-7}\,$m (see Fig.~\ref{fig1} for details). 
Also atom interferometry can be exploited to set bounds on the collapse parameters. To date, the most notable example is the experiment in Ref.~\cite{kovachy2015quantum}, where single atoms were placed in a superposition of the order of half-a-meter on a time-scale of a second. However, since these are single atoms ($m\sim 87\,\text{amu}$) the mass dependence strongly suppresses the advantages of the long times and large superposition distances.

With the same idea, one is also able to derive  a (reasonable) theoretical lower bound on the collapse parameters. Indeed, below such values the collapse mechanism loses its fundamental motivations: the model would allow for macroscopic superpositions to exist on a sufficiently long time-scale to be observed. Driven by this, one sets such a lower bound with the minimum values of the collapse parameters such that an object of dimensions $\sim 10\,\mu$m collapses in 0.01\,s, which are respectively the smallest visible size and the time resolution of the human eye.

\subsection{Non-interferometric tests}\label{non-int}

Contrary to interferometric experiments, non-interferometric tests do not rely on quantum superpositions. They exploit indirect effects of the action of collapse models, namely the presence of the collapse noise that imposes a jiggling on the system under scrutiny. As a consequence, one is able to exploit, beside microscopic systems, much larger and more massive systems, which can be well within the classical regime. Different experimental settings can thus exploit different effects to test or set upper bounds on the collapse parameters. We refer to Ref.~\cite{carlesso2022present} for a review on non-interferometric tests of collapse models and an overview of the relevant literature.

\begin{figure}[h!]
    \centering
    \includegraphics[width=0.8\linewidth]{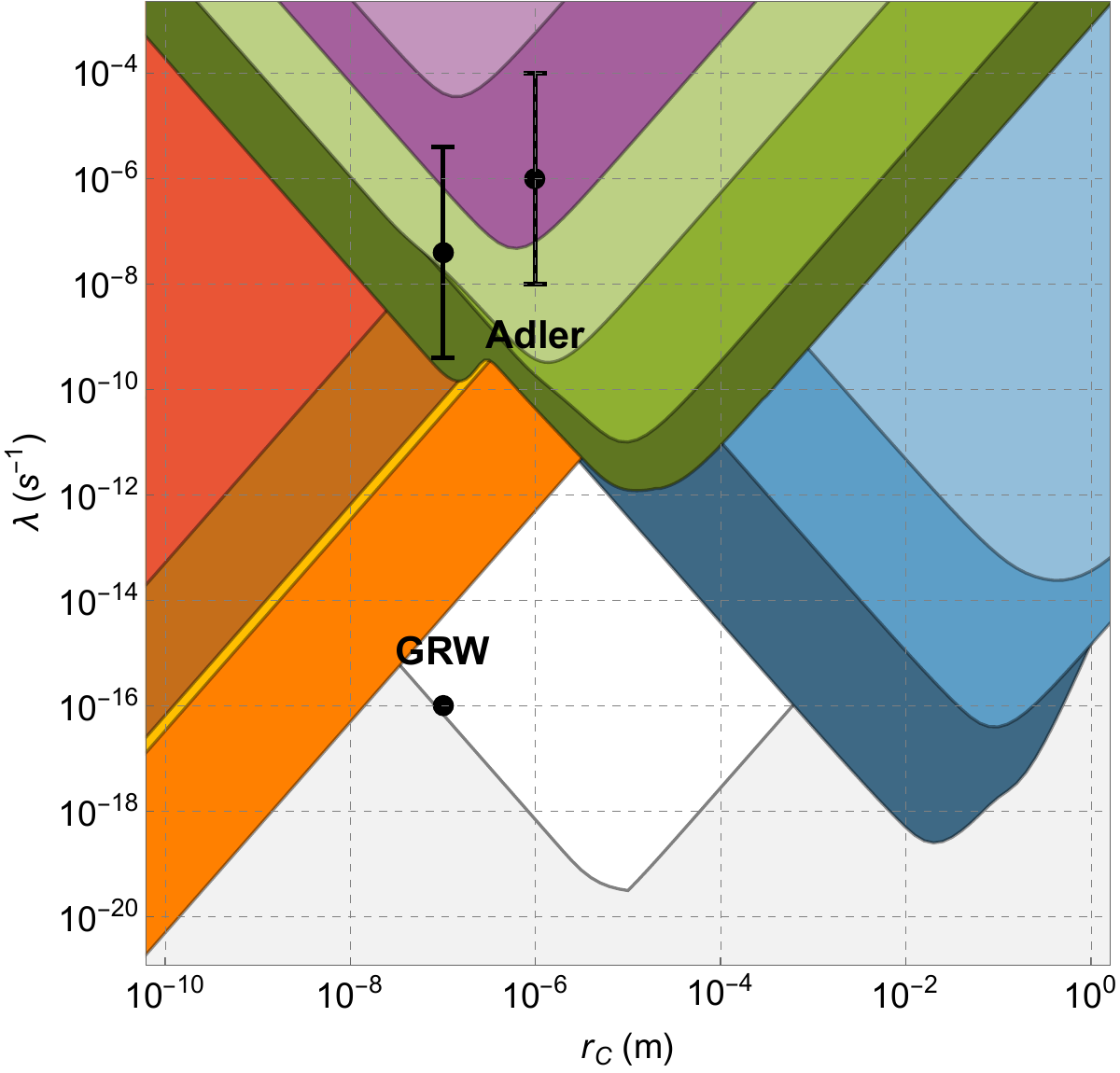}
    \caption{Exclusion plot for the CSL parameters $\lambda$ and $\rC$ from non-interferometric tests. The colored areas correspond to experimentally excluded regions. The yellow and brown areas are from phonon excitation in the CUORE experiment~\cite{adler2018bulk} and in Neptune~\cite{adler2019testing} respectively. The red area is from  {cold-atom experiment}~\cite{bilardello2016bounds}. The purple, green and blue areas are from different optomechanical experiments~\cite{pontin2020ultranarrow,zheng2020room,vinante2020narrowing,carlesso2016experimental,helou2017lisa}.
    The orange area is from spontaneous photon emission tests~\cite{arnquist2022search}.  The theoretical values proposed by GRW~\cite{ghirardi1986unified} and {the ranges proposed by} Adler~\cite{adler2007lower} are shown respectively as a black dot and black dots with bars that indicate the estimated range.  The theoretical values proposed by GRW\cite{ghirardi1986unified} and the ranges proposed
    by Adler\cite{adler2007lower} are shown respectively as a black dot and black dots
    with bars which indicate the estimated range. Finally, the light grey
    area is theoretically excluded\cite{torovs2017colored}. The white area has not been explored with non-interferometric experiments. Figure adapted from \cite{carlesso2022present}.
    }
    \label{fig:csl}
\end{figure}

\begin{figure}[h!]
    \centering
    \includegraphics[width=0.8\linewidth]{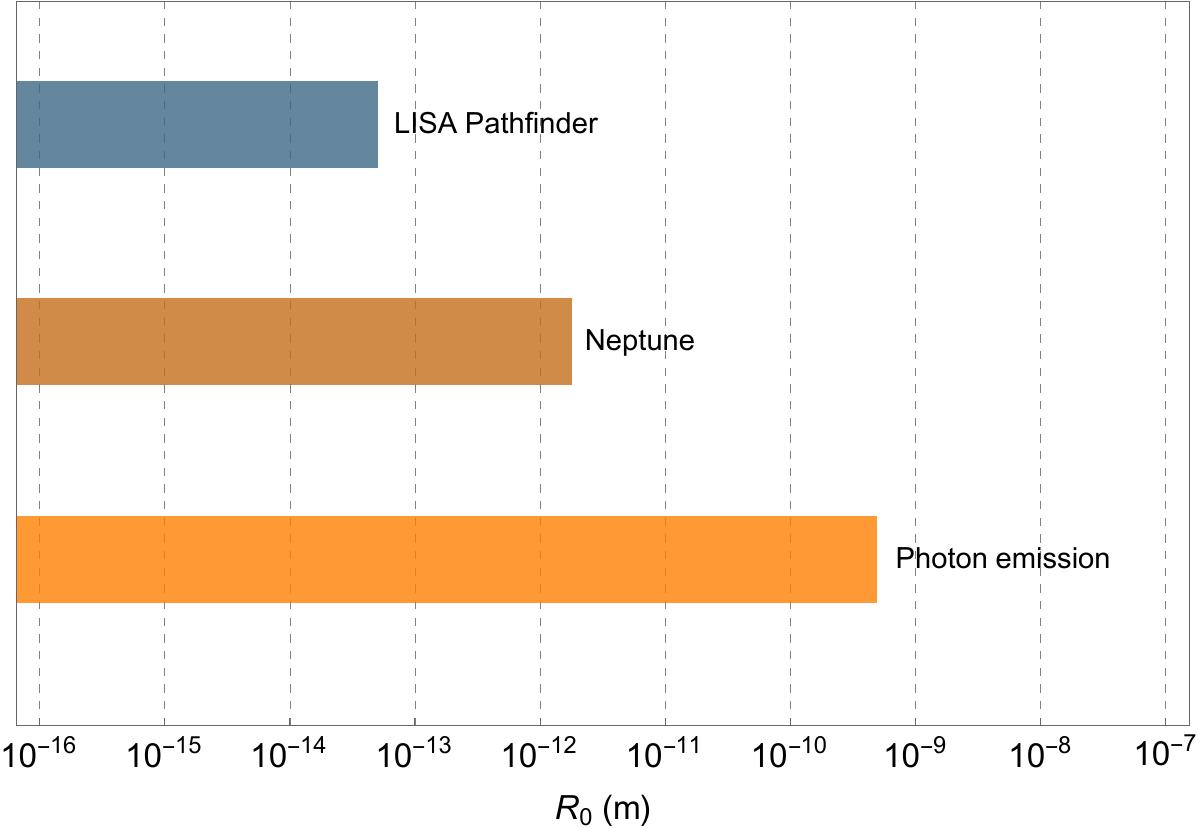}
    \caption{Exclusion plot for the DP parameter $R_0$ from non-interferometric tests. The colored areas correspond to experimentally excluded values of $R_0$. The brown area is from the heating rate of Neptune~\cite{adler2019testing}, the blue bound is from LISA Pathfinder~\cite{helou2017lisa}, the orange one is from radiation emission tests~\cite{arnquist2022search}. 
     Figure adapted from \cite{carlesso2022present}}
    \label{fig:dp}
\end{figure}

\subsubsection{Internal energy excitation}

The collapse noise induces an energy increase of all constituents of matter. In particular, it affects the collective dynamics of atoms (namely, phonons) in bulk materials, thus leading to an increase of the internal energy. The CSL model predicts a heating power $P_\text{\tiny CSL}=\tfrac{\D}{\D t}\braket{\hat H}$ given by
\begin{equation} \label{en_increase}
P_\text{\tiny CSL}=\frac{3}{4}\frac{\hbar^2 \lambda m}{m_0^2\rC^2},
\end{equation}
where $m$ is the total mass of the bulk system. To measure such an effect, one needs to isolate the system from other heating sources that would induce similar effects. Well isolated, low-temperature and high-vacuum experiments are relevant here. Underground facilities should be preferred to also isolate  from $\gamma$ radiation and cosmic rays. Under such conditions one can reach heating rates as low as  $P/m\sim 100$\,pW/kg. Accurate modelling of other energy depositions leads to an unaccounted heating power of $\sim 10$\,pW/kg, which sets the bound $\lambda<3.3\times10^{-11}$\,s$^{-1}$ at $\rC=10^{-7}$\,m for the CSL model~\cite{adler2018bulk}.

By exploiting the mass dependence in Eq.~\eqref{en_increase}, one can apply the same argument to massive astrophysical bodies. In detail, one equates the collapse-induced heating in Eq.~\eqref{en_increase} with the energy loss due to blackbody radiation emission (as described by the Stefan-Boltzmann law, $P_\text{rad}=S\sigma T^4$ where $S$ is the surface of the body, $\sigma$  Stefan's constant, and $T$ the temperature). It follows that low temperature astronomical systems can provide important bounds, e.g.~such an observation on Neptune leads to the bound $\lambda<6.6\times10^{-11}$\,s$^{-1}$ at $\rC=10^{-7}$\,m for the CSL model~\cite{adler2019testing}.

\subsubsection{Cold atoms}

Besides their use in interferometric experiments, cold atoms can also be employed for testing collapse models  in a non-interferometric setting. Indeed, one can compute the position spread evolution of a cold atom cloud when in free fall and compare it with the predictions of collapse models. Namely, the CSL model predicts that the position variance grows as
\begin{equation}\label{x2t3}
    \braket{\hat {\bf x}^2}_t=\braket{\hat {\bf x}^2}^\text{\tiny QM}_t+\frac{\lambda\hbar^2}{2m_0^2\rC^2}t^3,
\end{equation}
where $\braket{\hat {\bf x}^2}^\text{\tiny QM}_t$ is the standard quantum mechanical evolution.
Experiments performed at low temperatures would allow to suppress the growth of $\braket{\hat {\bf x}^2}^\text{\tiny QM}_t$, thus allowing for a better characterization of other diffusive mechanisms. The 
cooling to pK temperatures and subsequent release in free-fall of a cold atom cloud  set the bound $\lambda<5.1 \times 10^{-8}$\,s$^{-1}$ at $\rC=10^{-7}$\,m for the CSL model~\cite{bilardello2016bounds}.

\subsubsection{Optomechanical systems}

One can determine the action of collapse models by inferring the steady-state properties of optomechanical systems, which consist of an optical component shining on a mechanical resonator. The action of collapse models  is to impose an extra noise on the latter component, and thus raise its effective temperature with respect to the temperature $T$ of the surrounding environment. In general, one characterizes the action of all  noise acting on the mechanical resonator through the corresponding power spectral density, which reads~\cite{carlesso2022present}:
\begin{equation}
S_\text{\tiny PSD}(\omega)=S_\text{opto}(\omega)+\frac{\hbar \omega m \gamma_\text{m} \coth(\hbar \omega/2\kB T)+S_\text{\tiny CM}}{m^2[(\omega_\text{eff}^2-\omega^2)^2+\gamma_\text{eff}^2\omega^2]},
\end{equation}
where $S_\text{opto}(\omega)$ determines the optical influences; the second term is the contribution from the environment, which is proportional to the mechanical mass $m$ and its damping $\gamma_\text{m}$ and depends on the temperature $T$ and the effective frequency and damping $\omega_\text{eff}$ and $\gamma_\text{eff}$ respectively. Finally, the third term $S_\text{\tiny CM}=\hbar^2\eta$ quantifies the contribution of collapse models, which can be interpreted as a growth of the energy (i.e.~effective temperature) of the resonator (here, $\eta=\eta_{xx}$ is the diffusion constant due to CSL defined in Eq.~\eqref{eta}). Indeed, one has that $\braket{\hat x^2}\sim \int\D\omega S_\text{\tiny PSD}(\omega)\propto E+\Delta E_\text{\tiny CM}$, where $E$ is the standard energy of the system and $\Delta E$ that imposed by collapse models. 

Encompassing several orders of magnitude in mass, optomechanical systems provide a strong testbed for collapse models. Experiments with masses from pg (levitated in a Paul trap \cite{pontin2020ultranarrow}) to tons (gravitational wave detectors \cite{carlesso2016experimental}) were exploited, and to date provide strong bounds on the CSL model for small values of $\rC$ (down to $\lambda<2.0\times 10^{-10}$\,s$^{-1}$ at $\rC=10^{-7}$\,m, which covers completely the values suggested by Adler for the CSL parameters \cite{vinante2020narrowing}), while giving the strongest bounds to date for large values of $\rC>10^{-5}\,$m.

\subsubsection{Collapse-induced photon emission}

As already underlined, the action of  collapse noise is to impose a jiggling to a massive system, which thus undergoes  random accelerations. By exploiting the well-known effect that a charged particle emits radiation if accelerated, one can bound the effect of  collapse noise by measuring the spontaneous photon emission of a charged particle. The CSL model predicts that a single atom is characterized by the following radiation emission rate~\cite{Donadi:2021tq}
\begin{equation}\label{eq:ratefinalevero}
\left.\frac{\D\Gamma_\text{\tiny CSL}}{\D E}\right|_\text{atom}=\frac{\left(N_{A}^{2}+N_{A}\right)\lambda \hbar e^{2}}{4\pi^{2}\varepsilon_{0}m_{0}^{2}\rC^{2}c^{3}E},
\end{equation}
which is valid for $E\in[10, 10^5]$\,keV  corresponding to photon wavelengths larger than nuclear size but smaller than atomic one, and where  $N_{A}$ is the atomic number, and $E$ the energy of the emitted photons. A similar expression can also be found  for the DP model~\cite{donadi2020underground}. Experiments in underground facilities such as the Gran Sasso laboratories in Italy and Sanford Underground Research Facility in South Dakota pose strong bounds by characterising the radiation emission from Germanium samples (here, one needs to multiply Eq.~\eqref{eq:ratefinalevero} by the number of atoms present in the sample), which are cooled to cryogenic temperatures. Such experiments bound the CSL model to $\lambda<4.9 \times 10^{-15}$\,s$^{-1}$ at $\rC=10^{-7}$\,m and the DP model to $R_0\ge4.9 \times 10^{-10}$\,m
~\cite{arnquist2022search}.

\subsubsection{Astronomical and cosmological setting}

Owning the concept that collapse effects are mass proportional and grow in time,  astronomical and cosmological settings provide an interesting test arena due to their characteristic long times and large  masses. Several situations were considered, from the dissociation of cosmic hydrogen during the evolution of the universe to the distortion of the radiation spectrum of the Cosmic Microwave Background (CMB), from the heating of neutron stars to that of the intergalactic medium (see \cite{carlesso2022present} and references therein).

In the cosmological context, collapse models were used to justify the emergence of an effective cosmological constant and of cosmic structures. Collapses will also impact the spectrum of primordial perturbations, whose action can be quantified in the structure of the CMB \cite{carlesso2022present}. 
Here, observations indicated that CSL model can be ruled out for a specific choice of the relativistic collapse operator~\cite{martin2020cosmic}. Nevertheless, a different choice restores compatibility of CSL with cosmological observations~\cite{gundhi2021impact} underlining that the generalization to a relativistic setting must be tackled with care.

\section{Generalization}

The CSL and the DP models include in their dynamical equations noise white. However, in nature  it is impossible to find perfectly white noises, with a completely flat spectrum, without a cutoff or a resonance characterizing their action. Thus white collapse noise can be considered to be an idealization and one should state what happens when considering non-white generalizations. Moreover, collapse dynamics breaks the energy conservation of a system that is  otherwise isolated by heating the system to an infinite temperature. Although such heating has a rather long time-scale, and thus can be neglected in most situations, this constant increase of energy, never reaching thermalization, is something that one would like to {avoid, also} in a phenomenological model. Two families of generalizations of collapse models, the colored and dissipative ones, have been developed to resolve these limitations, and are discussed below.

\subsection{Colored generalizations}

The inclusion of colored features structurally changes the dynamical equation for the wavefunction, which reads \cite{bassi2003dynamical}
\begin{equation}
    \frac{\D \ket{\psi(t)}}{\D t}=\left[  -\frac{i}{\hbar}\hat H+\sqrt{\gamma}\sum_{i=1}^N\hat A_iw_i(t)-2\gamma\sum_{i,j=1}^N \hat A_i\int_0^t\D s\, D_{ij}(t,s)\frac{\delta}{\delta w_j(s)} \right] \ket{\psi(t)},
\end{equation}
where $\delta/\delta w_j(s)$ indicates the functional derivative with respect to the noise $w_j(s)$. 
However, since this equation is complicated to handle, one adopts a perturbative approach, which is well motivated by the smallness of the collapse effects.
For simplicity, we opt for the description based on the imaginary noise trick as in Eq.~\eqref{ss}.
Here the noises $w_i(t)$ are described in terms of 
\begin{equation}
\mathbb E[w_i(t)]=0,\quad\text{and}\quad\mathbb E[w_i(t)w_j(s)]=D_{ij}(t,s).
\end{equation}
where  $D_{ij}(t,s)$ contains the time correlation function of the noise, which defines the noise to be white (i.e., $D_{ij}(t,s)\propto\delta(t-s)$) or colored (i.e., $D_{ij}(t,s)\not\propto\delta(t-s)$). The corresponding master equation becomes
    \begin{equation}\label{ME_colored}
    \begin{aligned}
    \frac{\D\hat\rho(t)}{\D t}=-\frac{i}{\hbar}\left[\hat H,\hat\rho(t)\right]+\gamma\sum_{i,j=1}^{N}\int_0^t\D s\,D_{ij}(t,s)
\left(
\hat A_i\hat\rho(t)\hat A_j(s-t)\right.\\
\left.+\hat A_j(s-t)\hat \rho(t)\hat A_i-\hat A_i\hat A_j(s-t)\hat \rho(t)-\hat \rho(t)\hat A_j(s-t)\hat A_i
\right)   ,
\end{aligned}
    \end{equation}
where $\hat A_j(s-t)=e^{\tfrac i\hbar \hat H (s-t)}\hat A_je^{-\tfrac i\hbar \hat H (s-t)}$. This master equation is  the result of a first order expansion in the noise coupling $\gamma$.

In the specific case of the colored generalization of the CSL model, the collapse dynamics (up to the order $\lambda$) is given by Eq.~\eqref{CSL_gau} 
 with the white noise $w(\x,t)=\D \bar W_t(\x)/\D t$ substituted with its colored version, for which the following relations hold
\begin{equation}
    \mathbb E[w(\x,t)]=0,\quad\mathbb E[w(\x,t)w(\y,t)]=G(\x-\y)f(t-s),
\end{equation}
where $G(\x-\y)=e^{-{(\x-\y)^{2}}/{4\rC^{2}}}$ describes the spatial correlation, and $f(t-s)$  the time correlation function of the collapse noise.  In general, the latter will be a non-trivial (colored) function of time, whose Fourier transform $\tilde f(\omega)$ will stray from a constant spectrum. The standard white noise case is recovered by requiring that $f(t-s)=\delta(t-s)$. 

The effects of the colored modification of the noise can be accounted in the following. For the interferometric tests, the expression for $F_\text{\tiny CSL}(\k,\q,t)$ in Eq.~\eqref{eq.rho_free_CSL} becomes $F_\text{\tiny cCSL}$, where  \cite{torovs2017colored}
\begin{equation}
    F_\text{\tiny cCSL}(\k,\q,t)=F_\text{\tiny CSL}(\k,\q,t) \exp\left[  \frac{\lambda \bar \tau}{2}\left(e^{-\tfrac{(\q-\k t/m)^2)}{4\rC^2}}-e^{-\tfrac{\q^2}{4\rC^2}}
    \right) \right],
\end{equation}
and where $\bar \tau =\int_0^t\D s f(s) s$. For non-interferometric tests, a case-by-case study is needed. In the case of internal energy increase, {the heating power takes the same form as in the white noise case displayed} in Eq.~\eqref{en_increase} where $\lambda$ is replaced with \cite{carlesso2018colored}
\begin{equation}
   \lambda_\text{eff}=\frac{2\lambda \rC^5}{3 \pi^{3/2}}\int \D \q\, e^{-\q^2 \rC^2}\q^2 \tilde f(\omega_\text{\tiny L}(\q)), 
\end{equation}
and $\omega_\text{\tiny L}$ denotes the longitudinal frequency of the phonons of the system. For cold atoms, which are treated as free non-interacting particles, the position spread in Eq.~\eqref{x2t3} is modified to 
\begin{equation}
\braket{\hat {\bf x}^2}_t=\braket{\hat {\bf x}^2}^\text{\tiny QM}_t+\frac{3\lambda A^2 \hbar^2}{m^2 \rC^2}\int_0^t\D s\,(t+s)^2\int_0^s\D s_1f(s_1).
\end{equation}
For optomechanical systems, the collapse contribution $S_\text{\tiny CM}$ is mapped to
\begin{equation}
    S_\text{\tiny cCM}=S_\text{\tiny CM}\tilde f(\omega),
\end{equation}
which directly imprints the same frequency dependence of $\tilde f(\omega)$ on the collapse contribution to the power spectral density. In a very similar way, for the collapse-induced photon emission the corresponding rate changes from Eq.~\eqref{eq:ratefinalevero} to 
\begin{equation}
\left.\frac{\D\Gamma_\text{\tiny cCSL}}{\D E}\right|_\text{atom}=\frac{\left(N_{A}^{2}+N_{A}\right)\lambda \hbar e^{2}}{4\pi^{2}\varepsilon_{0}m_{0}^{2}\rC^{2}c^{3}E}\times \tilde f\left(\frac{E}{\hbar}\right)=\left.\frac{\D\Gamma_\text{\tiny CSL}}{\D E}\right|_\text{atom}\times \tilde f\left(\frac{E}{\hbar}\right).
\end{equation}

By choosing a specific form of the time-correlation function, the bounds on the model are correspondingly modified. To be quantitative, we consider explicitly the case
\begin{equation}\label{eq.fcorr}
    f(t-s)=\frac{\Omega_\text{\tiny C}}{2}e^{-\Omega_\text{\tiny C}|t-s|},\quad\text{i.e.,}\quad\tilde f(\omega)=\frac{\Omega_\text{\tiny C}^2}{\Omega_\text{\tiny C}^2+\omega^2},
\end{equation}
where $\Omega_\text{\tiny C}$ is the cut-off frequency of the collapse noise and $\tau_\text{\tiny C}=\Omega_\text{\tiny C}^{-1}$ is its corresponding correlation time. Note that we have introduced a new phenomenological parameter $\Omega_\text{\tiny C}$, whose different values will give  different quantitative predictions and thus changes in the bounds. This explicit example was studied in \cite{carlesso2018colored}, where the bounds change based on the particular value of $\Omega_\text{\tiny C}$. 
While bounds from interferometric experiments do not change significantly with respect to the colored generalizations, non-interferometric ones do. In particular, predictions depending on high energy/frequency are heavily weakened by the introduction of the cut-off. Namely, if the characteristic frequency of the experiment is higher than $\Omega_\text{\tiny C}$, then the corresponding bound is suppressed. If one assumes that the collapse noise has a cosmological origin, then different arguments suggest a value of $\Omega_\text{\tiny C}\sim 10^{12}\,$Hz \cite{carlesso2018colored}.
Thus, eventually, only bounds from mechanical experiments at low frequency and those from free-falling cold atoms remain essentially untouched.

\subsection{Dissipative generalizations}

To include dissipative effects in the collapse equation, one needs to modify the structure of the collapse operator. We directly focus on the dissipative version of the CSL collapse equation, which reads \cite{smirne2015dissipative}
\begin{equation}\label{eq.dCSL}
\begin{aligned}
    \D \ket{\psi(t)}=&\left[
-\frac{i}{\hbar}\hat H\D t +\frac{\sqrt{\gamma}}{m_0}\int\D\y\,\left(\hat L(\y)-r_t(\y)\right)\D W_t(\y)\right.\\
&\left.-\frac{\gamma}{2m_0^2}\int\D \y\,\left(
\hat L^\dag(\y)\hat L(\y)+r_t^2(\y)-2r_t(\y)\hat L(\y)
\right)   \D t
    \right]\ket{\psi(t)},
\end{aligned}
\end{equation}
where $\gamma=\lambda(4\pi \rC^2)^{3/2}$, $r_t(\y)=\braket{\psi(t)|\left(\hat L^\dag(\y)+\hat L(\y)\right)|\psi(t)}/2$ and $W_t(\y)$ is an ensemble of independent Wiener processes, one for each point in space.
By restricting to a sector of the Fock space with a fixed number $N$ of particles (for simplicity, all of mass $m$), one defines the collapse operator as
\begin{equation}
    \hat L_\text{\tiny dCSL}(\y)=\frac{m}{(2\pi\hbar)^3}\sum_{\alpha=1}^N\int\D\q
\,e^{\tfrac{i}{\hbar}\q\cdot(\hat \x_\alpha-\y)}\exp\left(-\frac{\rC^2}{2\hbar^2}\left|(1+\chi)\q+2\chi\hat \p_\alpha\right|^2\right),
\end{equation}
which is the modified non-Hermitian collapse operator. The latter includes a new phenomenological parameter $\chi$, and indicates that the action of the noise will suppress high momenta. Indeed the momenta $\q$ exchanged by the noise follow a Gaussian distribution peaked around $-2\hat \p \chi/(1+\chi)$. Correspondingly, the energy of the system is also suppressed, as can be shown for example by studying the mean kinetic energy $E(t)=\braket{\hat \p^2/(2m)}$ for a single particle:
\begin{equation}
    \frac{\D E(t)}{\D t}=\frac{3\hbar^2 \lambda m}{4(1+\chi)^5 \rC^2 m_0^2}-\frac{4\chi \lambda m^2}{(1+\chi)^5 m_0^2}E(t).
\end{equation}
Then, one can define an effective temperature $T_\text{\tiny CSL}$ at which the system will thermalise asymptotically $E(t\to\infty)=\kB T_\text{\tiny CSL}$,  which can be also interpreted as the temperature of the collapse noise field. Such a temperature can be found by setting the above expression for ${\D E(t)}/{\D t}=0$, thus leading to 
\begin{equation}\label{eq.T.CSL}
    T_\text{\tiny CSL}=\frac{\hbar^2}{8 \kB \rC^2 m \chi}.
\end{equation}
Correspondingly the experimental predictions are also modified with respect to the standard CSL model. For interferometric tests, the expression for $F_\text{\tiny CSL}$ in Eq.~\eqref{Finterf.CSL} becomes \cite{torovs2017colored}
\begin{equation}
    F_\text{\tiny dCSL}(\k,\q,t)=\exp\left[-\lambda\frac{m^2}{m_0^2}\left(t-e^{-\tfrac{\k^2 \rC^2 \chi^2}{\hbar^2}}\int_0^t\D s\,e^{-\tfrac{(\q -\k s/m)^2}{4\rC^2(1+\chi)^2}}\right)\right],
\end{equation}
which is valid in the limit of small $\chi$. 
Cold atoms will diffuse according to \cite{bilardello2016bounds}
\begin{equation}
\begin{aligned}
    \braket{\hat \x^2}_t=&\braket{\hat \x^2}_0+\frac{2(\braket{\hat \p^2}_0-\braket{\hat \p^2}_\text{as})}{m^2(B-C)}\left(\frac{1-e^{-C t}}{C}-\frac{1-e^{-Bt}}{B}\right)\\
   & +\left(\braket{\{\x,\p\}}_0-\frac{2\braket{\hat \p^2}_\text{as}}{m B}\right)\frac{1-e^{-Bt}}{mB}+\left(\frac{6\lambda A^2 \rC^2 \chi^2}{(1+\chi)^3}+\frac{2{\braket{\hat \p^2}_\text{as}}}{m^2B}\right)t,
\end{aligned}
\end{equation}
where $B=(1+\chi)C/2$, $C=4\chi \lambda A^2/(1+\chi)^5$ and $\braket{\hat \p^2}_\text{as}=3\hbar^2/(8\chi \rC^2)$.
For optomechanical systems, one derives -- thorough a non-trivial unravelling technique -- that the effects of dissipations are twofold \cite{nobakht2018unitary}. On one side, one has a modification of the power spectral density that reads
\begin{equation}
S_\text{\tiny dCM}(\omega)=S_\text{\tiny CM}[1+\chi^2m^2(\gamma^2+\omega^2)],
\end{equation}
thus becoming also frequency dependent. On the other hand, an extra dissipation occurs, which adds to the standard one, and it reads
\begin{equation}\label{eq.gamma-CSL}
\gamma_\text{\tiny CSL}=\eta_\text{\tiny dCSL} \frac{4\rC^2 m_0\chi(1+\chi)}{m},
\end{equation}
where $\eta_\text{\tiny dCSL}$ can be found from $\eta$ [cf.~Eq.~\eqref{eta}] by mapping $\rC\mapsto\rC (1+\chi)$. By making a similar assumption as for the colored generalization that the noise has a cosmological origin, {a reasonable value of $T_\text{\tiny CSL}$ can be set to 1\,K. This is of the order of magnitude of the temperature of the Cosmological Microwave Background.} One finds that for values of $T_\text{\tiny CSL}$  higher than 1\,K, no significant changes are shown by the bounds. Indeed, the value of $\chi$ [cf.~Eq.~\eqref{eq.T.CSL}] is too small to show deformations with respect to the standard CSL in the diffusion constant $\eta_\text{\tiny dCSL}$ or to give significant values for the extra dissipative constant $\gamma_\text{\tiny CSL}$ [cf.~Eq.~\eqref{eq.gamma-CSL}].
Conversely, for lower values of $T_\text{\tiny CSL}$, deformations of bounds appear and are stronger for smaller values of $\rC$. Bounds on the dissipative version of the CSL model have been set from interferometric experiments, which do not change with respect to those shown in Fig.~\ref{fig1} for the standard CSL, from cold-atom and optomechanical experiments.

The dissipative version of the DP model has a collapse equation, which is also given by Eq.~\eqref{eq.dCSL} where  $\gamma=4\pi\hbar m_0^2 G$ {and the collapse operator} is now \cite{bahrami2014role}
\begin{equation}
    \hat L_\text{\tiny dDP}(\y)=\frac{m}{(2\pi\hbar)^3}\sum_{\alpha=1}^N\int\D\q
\,\frac{e^{\tfrac{i}{\hbar}\q\cdot(\hat \x_\alpha-\y)}}{q}\exp\left(-\frac{R_0^2}{2\hbar^2}\left|(1+\chi)\q+2\chi\hat \p_\alpha\right|^2\right),
\end{equation}
and where the specific choice for $\chi=m_0/m$ is made. The corresponding expressions for the collapse noise temperature, the damping rate and the diffusion constant then read
\begin{equation}
\begin{aligned}
    T_\text{\tiny DP}=&\frac{\hbar^2}{8\kB m_0 R_0^2},\\
\gamma_\text{\tiny DP}=&4\eta_\text{\tiny DP} R_0^2\chi(1+\chi)\frac{m_0}{m},\\
\eta_\text{\tiny DP}=&\frac{Gm^2}{\sqrt{\pi}R^3}I\left(\frac{R}{R_0(1+\chi)}\right), \end{aligned}
\end{equation}
where $I(x)=\sqrt{\pi}\erf(x)+\tfrac{1}{x}(e^{-x^2}-3)+\tfrac{2}{x^2}(1-e^{-x^2})$.
To date, the dissipative DP model has been tested only  with optomechanical systems, where measurements of the damping rate $\gamma$ were performed. 
However, for the natural values of the parameters $R_0=10^{-15}$\,m and 
$T\simeq 1$\,K,
the model becomes inconsistent for systems with masses smaller than $10^{11}$\,amu, since it predicts dissipative effects that are experimentally excluded. Thus, either one admits that the dissipative DP model is an effective model only valid for mesoscopic systems or one considers larger values of $R_0$. By doing so, the model is consistent even for smaller masses, but one pays the price that $R_0$ becomes a free phenomenological parameter completely unrelated to the actual size of nucleons.

Finally, we underline that there is a certain freedom in choosing the collapse operator
for a dissipative collapse model. For example, in \cite{di2023linear} a simpler choice is considered for both the CSL and the DP models.
This can be introduced   by employing the structure in Eq.~\eqref{eq.dCSL} for the collapse equation of both models, with $\gamma=m_0^2/\hbar^2$, and having that the Fourier transform of the spatial correlations of the noise $D(\x-\y)$ reads
\begin{equation}
    \tilde D_\q=
    \begin{cases}
        \frac{4\pi\hbar G}{q^2} e^{-R_0^2 q^2},&\text{for the DP model,}\\
       \frac{ \lambda(4\pi \rC^2)^{3/2}\hbar^2}{m_0^2} e^{-\rC^2q^2},&\text{for the CSL model.}
    \end{cases}
\end{equation}
A new, dissipative collapse operator is  defined as 
\begin{equation}
    \hat L(\y)=\hat M(\y)-i\frac{\hbar \beta}{4}\nabla_\y \hat {\bf J}(\y),
\end{equation}
where $\hat M(\y)=m\hat \psi^\dag(\y)\hat\psi(\y)$ is the standard mass density operator (where $\hat \psi^\dag(\y)$ creates a particle of mass $m$ in the position $\y$) and
\begin{equation}
    \hat {\bf J}(\y)=-i\frac{\hbar}{2}\left(\hat \psi^\dag(\y)\nabla_\y\hat \psi(\y)-\nabla_\y\hat \psi^\dag(\y)\hat \psi(\y)\right),
\end{equation}
is the standard current. Here, $\beta$ is a new phenomenological parameter quantifying the dissipative strength of the model and thus the collapse noise field temperature. Since its expression is not as trivial as those given above, we refer the reader to \cite{di2023linear} for its explicit form for these versions of the dissipative extensions of the CSL and the DP models.

\section{Conclusion}

Collapse models provide a pragmatic, phenomenological solution to the quantum measurement problem. They are not just a different interpretation of quantum mechanics, but fundamentally change the rules of quantum theory. As with many modifications of quantum mechanics, they were always considered as restricted to theoretical investigations only. However in the last decade  technological advancements in precise manipulations and measurement of (quantum) systems has finally allowed for the development of the experimental endeavour. Such experiments not only bound the values of collapse model parameters, but also constitute a way of parameterizing the limits of the validity of the superposition principle.

\section*{Acknowledgments}

M.C. acknowledges the UK
EPSRC (Grant No.~EP/T028106/1), the EU EIC Pathfinder project QuCoM (10032223) and the PNRR PE National Quantum Science and Technology Institute (PE0000023). 
S.D. acknowledges support from the  Marie Sklodowska Curie Action through the UK Horizon Europe guarantee administered by UKRI. 

\bibliography{main.bib}{}

\begin{thebibliography}{10}

\bibitem{anderson2001science}
Philip~W Anderson.
\newblock Science: A ‘dappled world’ or a ‘seamless web’?, 2001.

\bibitem{carlesso2022present}
Matteo Carlesso, Sandro Donadi, Luca Ferialdi, Mauro Paternostro, Hendrik
  Ulbricht, and Angelo Bassi.
\newblock Present status and future challenges of non-interferometric tests of
  collapse models.
\newblock {\em Nature Physics}, 18(3):243--250, 2022.

\bibitem{bassi2003dynamical}
Angelo Bassi and GianCarlo Ghirardi.
\newblock Dynamical reduction models.
\newblock {\em Physics Reports}, 379(5-6):257--426, 2003.

\bibitem{gisin1989stochastic}
Nicolas Gisin.
\newblock Stochastic quantum dynamics and relativity.
\newblock {\em Helv. Phys. Acta}, 62(4):363--371, 1989.

\bibitem{ghirardi1986unified}
Gian~Carlo Ghirardi, Alberto Rimini, and Tullio Weber.
\newblock Unified dynamics for microscopic and macroscopic systems.
\newblock {\em Physical Review D}, 34(2):470, 1986.

\bibitem{ghirardi1990markov}
Gian~Carlo Ghirardi, Philip Pearle, and Alberto Rimini.
\newblock Markov processes in hilbert space and continuous spontaneous
  localization of systems of identical particles.
\newblock {\em Physical Review A}, 42(1):78, 1990.

\bibitem{adler2007collapse}
Stephen~L Adler and Angelo Bassi.
\newblock Collapse models with non-white noises.
\newblock {\em Journal of Physics A: Mathematical and Theoretical},
  40(50):15083, 2007.

\bibitem{donadi2022seven}
Sandro Donadi and Angelo Bassi.
\newblock Seven nonstandard models coupling quantum matter and gravity.
\newblock {\em AVS Quantum Science}, 4(2):025601, 2022.

\bibitem{diosi1989models}
Lajos Di{\'o}si.
\newblock Models for universal reduction of macroscopic quantum fluctuations.
\newblock {\em Physical Review A}, 40(3):1165, 1989.

\bibitem{penrose1996gravity}
Roger Penrose.
\newblock On gravity's role in quantum state reduction.
\newblock {\em General relativity and gravitation}, 28:581--600, 1996.

\bibitem{donadi2020underground}
Sandro Donadi, Kristian Piscicchia, Catalina Curceanu, Lajos Di{\'o}si,
  Matthias Laubenstein, and Angelo Bassi.
\newblock Underground test of gravity-related wave function collapse.
\newblock {\em Nature Physics}, 17:74--78, 2021.

\bibitem{arnquist2022search}
IJ~Arnquist, FT~Avignone~III, AS~Barabash, CJ~Barton, KH~Bhimani, E~Blalock,
  B~Bos, M~Busch, M~Buuck, TS~Caldwell, et~al.
\newblock Search for spontaneous radiation from wave function collapse in the
  majorana demonstrator.
\newblock {\em Physical Review Letters}, 129(8):080401, 2022.

\bibitem{torovs2017colored}
Marko Toro{\v{s}}, Giulio Gasbarri, and Angelo Bassi.
\newblock Colored and dissipative continuous spontaneous localization model and
  bounds from matter-wave interferometry.
\newblock {\em Physics Letters A}, 381(47):3921--3927, 2017.

\bibitem{kovachy2015quantum}
T~Kovachy, P~Asenbaum, C~Overstreet, CA~Donnelly, SM~Dickerson, A~Sugarbaker,
  JM~Hogan, and MA~Kasevich.
\newblock Quantum superposition at the half-metre scale.
\newblock {\em Nature}, 528(7583):530--533, 2015.

\bibitem{adler2007lower}
Stephen~L Adler.
\newblock Lower and upper bounds on csl parameters from latent image formation
  and igm heating.
\newblock {\em Journal of Physics A}, 40(12):2935, 2007.

\bibitem{adler2018bulk}
Stephen~L Adler and Andrea Vinante.
\newblock Bulk heating effects as tests for collapse models.
\newblock {\em Physical Review A}, 97(5):052119, 2018.

\bibitem{adler2019testing}
Stephen~L Adler, Angelo Bassi, Matteo Carlesso, and Andrea Vinante.
\newblock Testing continuous spontaneous localization with fermi liquids.
\newblock {\em Physical Review D}, 99(10):103001, 2019.

\bibitem{bilardello2016bounds}
Marco Bilardello, Sandro Donadi, Andrea Vinante, and Angelo Bassi.
\newblock Bounds on collapse models from cold-atom experiments.
\newblock {\em Physica A}, 462:764--782, 2016.

\bibitem{pontin2020ultranarrow}
A~Pontin, NP~Bullier, M~Toro{\v{s}}, and PF~Barker.
\newblock Ultranarrow-linewidth levitated nano-oscillator for testing
  dissipative wave-function collapse.
\newblock {\em Physical Review Research}, 2(2):023349, 2020.

\bibitem{zheng2020room}
Di~Zheng, Yingchun Leng, Xi~Kong, Rui Li, Zizhe Wang, Xiaohui Luo, Jie Zhao,
  Chang-Kui Duan, Pu~Huang, Jiangfeng Du, et~al.
\newblock Room temperature test of the continuous spontaneous localization
  model using a levitated micro-oscillator.
\newblock {\em Physical Review Research}, 2(1):013057, 2020.

\bibitem{vinante2020narrowing}
A~Vinante, M~Carlesso, A~Bassi, A~Chiasera, S~Varas, P~Falferi, B~Margesin,
  R~Mezzena, and H~Ulbricht.
\newblock Narrowing the parameter space of collapse models with ultracold
  layered force sensors.
\newblock {\em Physical Review Letters}, 125(10):100404, 2020.

\bibitem{carlesso2016experimental}
Matteo Carlesso, Angelo Bassi, Paolo Falferi, and Andrea Vinante.
\newblock Experimental bounds on collapse models from gravitational wave
  detectors.
\newblock {\em Physical Review D}, 94(12):124036, 2016.

\bibitem{helou2017lisa}
Bassam Helou, BJJ Slagmolen, David~E McClelland, and Yanbei Chen.
\newblock Lisa pathfinder appreciably constrains collapse models.
\newblock {\em Physical Review D}, 95(8):084054, 2017.

\bibitem{Donadi:2021tq}
Sandro Donadi, Kristian Piscicchia, Raffaele~Del Grande, Catalina Curceanu,
  Matthias Laubenstein, and Angelo Bassi.
\newblock {Novel CSL bounds from the noise-induced radiation emission from
  atoms}.
\newblock {\em The European Physical Journal C}, 81:773, 2021.

\bibitem{martin2020cosmic}
J{\'e}r{\^o}me Martin and Vincent Vennin.
\newblock Cosmic microwave background constraints cast a shadow on continuous
  spontaneous localization models.
\newblock {\em Physical Review Letters}, 124(8):080402, 2020.

\bibitem{gundhi2021impact}
Anirudh Gundhi, Jos{\'e}~Luis Gaona-Reyes, Matteo Carlesso, and Angelo Bassi.
\newblock Impact of dynamical collapse models on inflationary cosmology.
\newblock {\em Phys. Rev. Lett.}, 127:091302, 2021.

\bibitem{carlesso2018colored}
Matteo Carlesso, Luca Ferialdi, and Angelo Bassi.
\newblock Colored collapse models from the non-interferometric perspective.
\newblock {\em The European Physical Journal D}, 72:1--7, 2018.

\bibitem{smirne2015dissipative}
Andrea Smirne and Angelo Bassi.
\newblock Dissipative continuous spontaneous localization (csl) model.
\newblock {\em Scientific reports}, 5(1):1--9, 2015.

\bibitem{nobakht2018unitary}
Jahangir Nobakht, Matteo Carlesso, Sandro Donadi, Mauro Paternostro, and Angelo
  Bassi.
\newblock Unitary unraveling for the dissipative continuous spontaneous
  localization model: Application to optomechanical experiments.
\newblock {\em Physical Review A}, 98(4):042109, 2018.

\bibitem{bahrami2014role}
Mohammad Bahrami, Andrea Smirne, and Angelo Bassi.
\newblock Role of gravity in the collapse of a wave function: A probe into the
  di{\'o}si-penrose model.
\newblock {\em Physical Review A}, 90(6):062105, 2014.

\bibitem{di2023linear}
Giovanni Di~Bartolomeo, Matteo Carlesso, Kristian Piscicchia, Catalina
  Curceanu, Maaneli Derakhshani, and Lajos Di{\'o}si.
\newblock On the linear friction many-body equation for dissipative spontaneous
  wavefunction collapse.
\newblock {\em arXiv preprint arXiv:2301.07661; Accepted in Physical Review A},
  2023.

\end{thebibliography}
\bibliographystyle{unsrt}

\end{document}